\documentclass[%
superscriptaddress,
preprint,
 amsmath,amssymb,
 aps,
prc
]{revtex4-1}

\usepackage{color}
\usepackage{graphicx}
\usepackage{hyperref}

\begin{document}

\title{Active gel model for one-dimensional cell migration coupling actin flow and adhesion dynamics}
\author{Valentin W{\"o}ssner}
\affiliation{Institute for Theoretical Physics, Heidelberg University, Germany}
\affiliation{BioQuant, Heidelberg University, Germany}
\author{Oliver M. Drozdowski}
\affiliation{Institute for Theoretical Physics, Heidelberg University, Germany}
\affiliation{BioQuant, Heidelberg University, Germany}
\affiliation{Max Planck School Matter to Life, Heidelberg, Germany}
\author{Falko Ziebert}
\affiliation{Institute for Theoretical Physics, Heidelberg University, Germany}
\affiliation{BioQuant, Heidelberg University, Germany}
\author{Ulrich S.  Schwarz}
\affiliation{Institute for Theoretical Physics, Heidelberg University, Germany}
\affiliation{BioQuant, Heidelberg University, Germany}
\affiliation{Max Planck School Matter to Life, Heidelberg, Germany}
\date{\today}

\begin{abstract}
Migration of animal cells is based on the interplay
between actin polymerization at the front,
adhesion along the cell-substrate interface,
and actomyosin contractility at the back.
Active gel theory has been used before 
to demonstrate that actomyosin contractility is sufficient for 
polarization and self-sustained cell migration in the absence of external cues, 
but did not consider the dynamics
of adhesion. Likewise, migration models based on
the mechanosensitive dynamics of adhesion receptors
usually do not include the global dynamics of intracellular flow.
Here we show that both aspects can be combined
in a minimal active gel model for one-dimensional cell migration with dynamic adhesion. 
This model demonstrates that load
sharing between the adhesion receptors leads to
symmetry breaking, with stronger adhesion at the
front, and that bistability of migration
arises for intermediate adhesiveness. 
Local variations in adhesiveness are sufficient 
to switch between sessile and motile states, in qualitative
agreement with experiments.
\end{abstract}

\maketitle

\section{Introduction}

In multicellular organisms like ourselves, each and every
cell has the ability to actively move \cite{bodor_cell_2020}. While cell migration of all cells plays an essential role during development, most cell types become quiescent in the 
mature organism, except for special situations like
wound healing and immune surveillance. 
Reawakening of the ability of locomotion enables metastatic tumor cells to spread throughout the body \cite{stuelten_cell_2018}. Single-cell motility 
relies on the interplay between many proteins, 
but the main ones are filamentous actin, the motor protein 
non-muscle myosin II, and the adhesion receptors of the integrin-family. 
The main cellular processes contributing to cell migration are 
the formation of actin-driven protrusion at the leading edge, force transmission onto the substrate mediated by integrin-based cell-matrix adhesion, and retraction of the rear by actomyosin contraction \cite{abercrombie_croonian_1980}. 
It is a striking observation that single cells
can switch between sessile and motile states
and also between different directions of migration \cite{verkhovsky_self-polarization_1999,Ziebert2012,lavi_deterministic_2016, ron_one-dimensional_2020, amiri_multistability_2023}.
To explain this bistable migration behaviour in the absence of external cues, one has to identify the underlying mechanisms for spontaneous symmetry breaking (SSB).

To simplify both experimental observation and theoretical modeling, cells can be placed on one-dimensional (1D) tracks \cite{maiuri_first_2012,maiuri_actin_2015,hennig_stick-slip_2020, schreiber_adhesionvelocity_2021, amiri_multistability_2023}, reducing the effective dimensionality of the problem. Such 
1D tracks can be made e.g. with microcontact
printing, laser lithography or microfluidics.
A natural framework to describe
cell migration in 1D is active gel theory \cite{kruse_generic_2005,julicher_active_2007,prost_active_2015}. 
This model class can represent both polymerization and
contractility in one coherent mathematical framework.
In particular, active gel theory has been used to
demonstrate that actomyosin contractility can be
a mechanism for SSB
that leads to self-sustrained cell motility, namely through
an instability during which myosin 
localizes to the rear of the cell \cite{recho_contraction-driven_2013}. Due to the spatial nature of the model, intracellular flows and concentration fields are accessible and can be compared to experiments \cite{recho_mechanics_2015}. Furthermore, the effects of external perturbations, such as optogenetics \cite{wittmann_lights_2020}, can be incorporated into active gel models as local
perturbations \cite{drozdowski_optogenetic_2021, drozdowski_optogenetic_2023}, again allowing for a comparison of experiment and theory.

An alternative mechanisms for SSB is the dynamics of adhesion. 
Adhesion receptors like the integrins are known to bind and rupture with mechanosensitive rates \cite{humphrey_mechanotransduction_2014}. 
It is well known that non-linear adhesion rates lead
to bifurcations in the adhesion dynamics \cite{bell_models_1978,erdmann_stability_2004},
which in turn lead to non-trivial adhesion profiles
and stick-slip motion of cell adhesion \cite{sabass_modeling_2010,schwarz_physics_2013}.
In general, stick-slip motion is typical for sliding friction
with non-linear dynamics \cite{schallamach_theory_1963,filippov_friction_2004}. 
Several bond models have been suggested to explain cell movement on 1D lines \cite{sens_stickslip_2020, ron_one-dimensional_2020, amiri_multistability_2023}. 
Such models focus on the left and right cell edges,
where they consider local force balance and retrograde flow, but they do not represent intracellular actin flow in a spatially resolved manner.
The interplay between
actin flow and adhesion dynamics is usually present in molecular clutch models \cite{chan_traction_2008,bangasser_determinants_2013,oria_force_2017,elosegui-artola_control_2018,alonso-matilla_optimal_2023},
but this model class usually assumes that SSB has already occured and in general
combines many individual rules to a complicated simulation framework.

Here we aim at representing both intracellular flow
and adhesion dynamics in a transparent active gel model that allows us to perform
a comprehensive mathematical analysis, using the powerful tools available for partial
differential equations. Recently, a somewhat similar approach has been presented by extending
the myosin-driven active gel model by adhesion dynamics, leading
to oscillatory states and stick-slip migration \cite{lo_mechanosensitive_2024}.
However, in this model adhesion played only a minor
role, while in practise, it might play
a dominant role for SSB. Experimentally, it is not clear
if one process dominates over the other. 
While some studies indicate that the cell front forms first via enhanced actin polymerization \cite{pollard_cellular_2003, ridley_cell_2003}, others suggest an increase in rear contractility as main mechanism \cite{cramer_forming_2010, yam_actinmyosin_2007,barnhart_balance_2015}. In both cases, the onset of polarization is assigned to the actomyosin system. In contrast, Hennig et al. hypothesized that a spontaneous loss of adhesion at one cell edge is responsible for the observed sudden decrease in traction of cells on 1D tracks \cite{hennig_stick-slip_2020}. This in turn causes the retraction of the prospective rear and initiates migration without pre-established cytoskeletal polarity.

To investigate the role of adhesion dynamics in the context
of contractility-based cell migration, here we propose a minimal 1D active gel model in which an explicit adhesion field is coupled to the intracellular flow of actin
in similar manner as in previously suggested bond models \cite{ron_one-dimensional_2020, sens_stickslip_2020},
but now in a spatially resolved fashion.
The adhesion density is treated as a reaction-diffusion system, where mechanosensitive bonds are subject to load sharing, which is known to be an essential non-linear effect leading to
stick-slip in cell adhesion \cite{bell_models_1978,schwarz_physics_2013}. In combination with symmetric edge polymerization, this approach constitutes a minimal model for a novel and simple motility mechanism for 1D cell migration, where SSB results from the adhesion dynamics. For intermediate adhesiveness, robust motility exists in a bistable regime and our model predicts adhesion and intracellular flow profiles in very good agreement with experimental observations. 

Having established this mechanism in our 1D-framework allows us to numerically study motility initiation by applying local perturbations to the adhesion density, similar to what has been demonstrated before for optogenetics \cite{drozdowski_optogenetic_2021,drozdowski_optogenetic_2023}. We show that nonlinear perturbations are required to switch from sessile to motile states and, in doing so, demonstrate the potential fundamental role of adhesion by forming the prospective rear. Furthermore, the interaction with a structured environment, e.g. a spatially varying concentration of ligands in the substrate, can be incorporated. In general, the model correctly recapitulates haptotactic behaviour \cite{ricoult_substrate-bound_2015} and we describe its predictions for cells on patterned lines \cite{schreiber_adhesionvelocity_2021, amiri_multistability_2023}.

This work is structured as follows.
In section \ref{sec:Minimal model} we introduce our model, derive consistent boundary conditions for the adhesion density and formulate a dimensionless boundary value problem (BVP). We then explain step-by-step the underlying polarization mechanism and how polarity is converted into directed migration by polymerization-driven flow. We show that active forces and adhesion-mediated friction have to be balanced and that motility is only possible in an intermediate regime of adhesiveness. In section \ref{sec:Induced failure of adhesion bonds} we analyse the effect of external perturbations, and in section \ref{sec:patterned_adhesion} we conduct a numerical study of cells on patterned lines to explore the full range of behaviour predicted by our model.

\section{Minimal model for adhesion-based motility} \label{sec:Minimal model}

\subsection{Model definition} \label{sec:Model definition}

\begin{figure}[t]
    \centering
    \includegraphics[width=15cm,keepaspectratio]{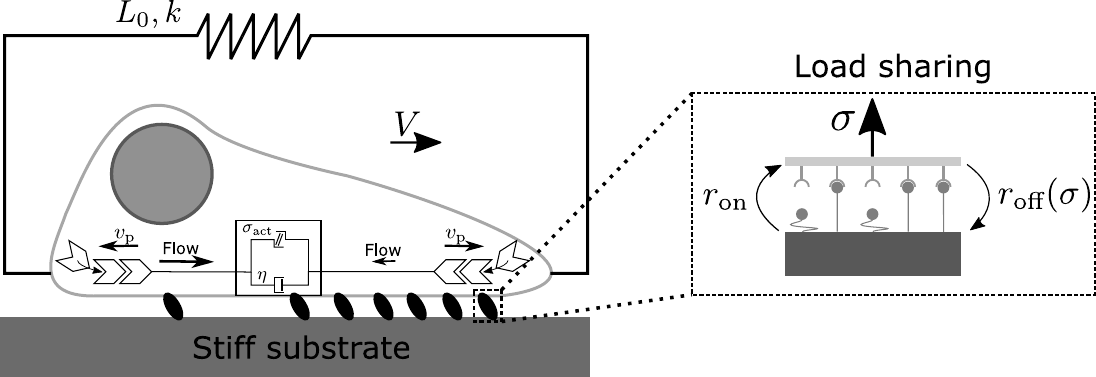}
    \caption[]{Schematic sketch of the 1D model: actomyosin contraction and edge polymerization of actin drive retrograde flow of a viscous bulk. Substrate friction is created by adhesion complexes, which are formed by binding/unbinding of mechanosensitive bonds, subject to load sharing, cf.~the zoomed-in sketch. The cell length is a dynamic quantity and couples the two edges elastically.}
    \label{fig:Schematic model}
\end{figure}

In the following we aim to identify the minimal set of
assumptions required  to model
the interplay between intracellular flow and adhesion
during 1D cell migration. Fig.~\ref{fig:Schematic model} sketches
the main elements. 
We model the passive response of the cytoskeleton as a purely viscous, infinitely compressible fluid, which is
locally subjected to an active stress $\sigma_{\rm{act}}$, modeling the effect of actomyosin contraction.
Since here we focus on the role of adhesion, 
$\sigma_{\rm{act}}$ is taken to be a constant, 
although earlier work also considered spatially
resolved contractility \cite{recho_contraction-driven_2013,lo_mechanosensitive_2024}. 
The constitutive stress equation then reads
\begin{equation} \label{eq:constitutive stress}
    \sigma(x,t) = \eta \partial_x v(x,t) + \sigma_{\rm{act}},
\end{equation}
with the viscosity $\eta$ and the actin flow velocity $v(x,t)$. In the molecular clutch model, the actin flow is coupled to the external substrate by receptor-ligand bonds mediated by proteins of the integrin-family, slowing down the retrograde flow and allowing for force transmission onto the substrate \cite{chan_traction_2008,giannone_multi-level_2009,oria_force_2017}. Assuming a linear relation between friction and local density of closed adhesion bonds $a(x,t)$ \cite{tawada1991protein, sabass_modeling_2010, ron_one-dimensional_2020, schreiber_adhesionvelocity_2021}, local force balance implies
\begin{equation} \label{eq:force balance}
    \partial_x \sigma(x,t) = \xi v(x,t) \left(a_0 + a(x,t) \right).
\end{equation}
$\xi$ is an effective friction coefficient and $a_0$ represents a baseline level of friction caused by other, non-specific dissipative interactions between actin, membrane and substrate. Combining equations Eq.~\ref{eq:constitutive stress} and Eq.~\ref{eq:force balance} yields a single stress equation
\begin{equation}
    \frac{\eta}{\xi} \partial_x \left(\frac{\partial_x \sigma(x,t)}{a_0 + a(x,t)}\right) = \sigma(x,t) - \sigma_{\rm{act}}.
\end{equation}

Single integrin bonds cluster to focal complexes and mature into so-called focal adhesions by recruiting further adapter proteins. We use a reaction-diffusion system \cite{barnhart_adhesion-dependent_2011,LoeberSM} to describe the effective evolution of the adhesion density
\begin{equation} \label{eq:Adhesion evolution}
    \partial_t a(x,t) = r_{\rm{on}} - r_{\rm{off}}(x,t) a(x,t) + D \partial_x^2 a(x,t).
\end{equation}
In a coarse-grained manner, we assume a dense distribution of ligands on the substrate and describe adhesion as a continuous field. 
The integrin receptors are abundant in the plasma membrane and in addition can switch between inactive and active states \cite{spiess_active_2018}, thus allowing for a dynamic recruitment of bound adhesion molecules. 
They are assumed to attach with a constant association rate $r_{\rm{on}}$, while the dissociation rate $r_{\rm{off}}$ is mechanosensitive and increases exponentially under load \cite{bell_models_1978}. This choice has recently found
a rigorous foundation in a statistical model 
for receptor-ligand binding at the cell membrane
\cite{janes_first-principle_2022} and is consistent
with experiments on membrane binding \cite{monzel2015measuring}. The constant association rate
then results if one averages over the membrane fluctuations.
Using the principle of load sharing, the local stress is distributed among the bonds at each position, 
corresponding to the force per bond typically used in models for adhesion sites \cite{bell_models_1978,erdmann_stability_2004,erdmann_stochastic_2004, sens_stickslip_2020}, such that 
\begin{equation} \label{eq:off rate}
    r_{\rm{off}}(x,t) = r_0 \ {\rm{exp}} \left(\frac{|\sigma(x,t)|}{f_0 \left(a_0 + a(x,t) \right)} \right).
\end{equation}
A small decrease in $a$ increases the force per bond on the remaining ones, which in turn facilitates detachment and can result in a rupture cascade \cite{erdmann_stability_2004, erdmann_stochastic_2004}.
The parameter $a_0$ in the exponent is introduced to avoid a divergence of the off-rate for very small adhesion densities. In principle, its value could be different from the baseline adhesion level introduced in equation Eq.~\ref{eq:force balance}. However, we have checked that the precise value of $a_0$ in equation Eq.~\ref{eq:force balance} has only a small effect and does not change any results qualitatively, while $a_0$ has a more crucial role in equation Eq.~\ref{eq:off rate} (cf. section \ref{sec: polarization mechanism}). Since one would expect that a baseline level of friction affecting the actin flow would physically also buffer the force experienced by the adhesive bonds to some degree and to reduce the number of free parameters in our model, we set the two to be equal.
$f_0$ represents the characteristic force scale at which single bonds tend to rupture. The off-rate without load is given by $r_0$.

Because even clustered integrins in adhesions are subject to thermal movement, we include a diffusive term with diffusion constant $D$ \cite{wiseman2004spatial}. It furthermore stabilizes the system \cite{barnhart_adhesion-dependent_2011, LoeberSM} and ensures continuity of the adhesion profiles.
However, the change in adhesion density due to binding should be dominant, while diffusion should only be relevant on long time scales. For a value of $10^{-1} \mu \rm{m}^2/s$, chosen by us, the change in adhesion on the length scale of cells ($\sim 10 \mu \rm{m}$) is around $D/(10\mu\rm{m}^2)^2 = 10^{-3}$/s. Binding is of the order of 0.1/s (cf. \ref{sec:Parameter estimates}) and is thus the more frequent process. Experimental measurements suggest an even smaller diffusivity of around $10^{-3} \mu \rm{m}^2/s$ \cite{wiseman2004spatial} such that diffusion does not impair the predictions by our model.
Since adhesions stay stationary relative to the substrate, while the migrating cell moves over them \cite{zaidel-bar_hierarchical_2004}, advection does not occur in the lab frame of reference.

To close the equations for stress and adhesion density, we have to formulate appropriate boundary conditions (BCs). Since we deal with a moving cell, the positions of the right and left boundary $l_\pm(t)$, corresponding to the cell edges, as well as the cell length, $L(t) = l_+(t) - l_-(t)$, will vary over time. However, due to volume regulation of cells, 
the typical cell size should not vary much. 
We therefore apply symmetric and elastic stress BCs \cite{recho_contraction-driven_2013, drozdowski_optogenetic_2021}
\begin{equation}
    \sigma (l_\pm,t) = -k \frac{L(t) - L_0}{L_0},
\end{equation}
with effective spring constant $k$ and reference length $L_0$,
which both result from volume regulation. 

Actin predominantly polymerizes in the vicinity of the plasma membrane, i.e. at the left and right boundaries in our model. Non-motile cells show symmetric polymerization that is matched by retrograde flow. In motile cells, polymerization induces membrane protrusions at the leading edge, while depolymerization is observed in the bulk and rear of the cell. Since our effective 1D description is an average over the height and width of the lamellipodium, we assume the same constant polymerization velocity $v_{\rm{p}}$ pointing outward at the two boundaries. Bulk depolymerization, which is necessary for conservation of total actin mass, is neglected \cite{julicher_active_2007} in agreement with the assumptions of an infinitely compressible gel. Then, polymerization offsets the movement of the cell edges and the flow velocity of the gel like
\begin{equation}
    \dot{l}_\pm(t) = v(l_\pm,t) \pm v_{\rm{p}}.
\end{equation}

Finally, starting from a system of reaction-diffusion equations for bound and unbound adhesion sites and assuming conservation of the total number of binding sites, we can derive BCs for $a(x,t)$ by taking the limit of an infinite reservoir of unbound sites (cf. \ref{sec:Conservation adhesion sites}) and obtain
\begin{equation}
    \partial_x a(l_\pm,t) = - \frac{\dot{l}_\pm(t)}{D} a (l_\pm,t).
\end{equation}
Because closed bonds are not advected, but only diffuse, the ratio of edge movement and diffusion arises here.

\subsection{Non-dimensionalisation}

We non-dimensionalize the equations by rescaling length by $L_0$, time by
the inverse off-rate without load $1/r_0$ , stress by the effective spring constant $k$, and adhesion density by $k/(L_0^2 \xi r_0)$. By transforming to the internal coordinate of the cell, $u = (x - l_-)/L$, we can map the moving BVP to the unit interval with stationary boundaries. Then, the global movement of the cell is absorbed in the advection velocity field $\tilde{v} = \dot{G}/L + \dot{L}/L(u- 1/2)$ with the cell's center position $G = (l_+ + l_-)/2$. The full nondimensional BVP reads
\begin{eqnarray}
    \label{eq:nondim BVP stress}
    \frac{\mathcal{L}^2}{L^2} \partial_u \frac{\partial_u \sigma(u,t)}{\mathcal{A} + a(u,t)} = \sigma(u,t) - \sigma_{\rm{act}}, \\
    \label{eq:nondim BVP stress BCs}
    \sigma(u_\pm,t) = -(L(t) - 1), \\
    \label{eq:nondim BVP flow BCs}
    \dot{l}_\pm (t) = \frac{1}{L} \frac{\partial_u \sigma (u_\pm,t)}{\mathcal{A} + a(u_\pm,t)} \pm v_{\rm{p}}, \\
    \label{eq:nondim BVP adhesion}
    \partial_t a(u,t) - \tilde{v}(u,t) \partial_u a(u,t) = \mathcal{R}  - {\rm{exp}}{\left(\frac{1}{\mathcal{F}} \frac{|\sigma(u,t)|}{\mathcal{A} + a(u,t)} \right)} a(u,t) + \frac{\mathcal{D}}{L^2} \partial_u^2 a(u,t), \\
    \label{eq:nondim BVP adhesion BCs}
    \partial_u a(u_\pm,t) = - \frac{L a(u_\pm,t)}{\mathcal{D}} \dot{l}_\pm(t). 
\end{eqnarray}
In addition to $\sigma_{\rm{act}}$ and $v_{\rm{p}}$, our model is described by five dimensionless parameters: $\mathcal{L} =  \sqrt{\eta r_0/k}$ describes the viscous time scale relative to the binding time. 
$\mathcal{R} = \xi L_0^2 r_{\rm{on}}/k$ determines the binding rate. $\mathcal{F} = f_0/(\xi L_0^2 r_0)$ represent the typical rupture force of closed bonds and $\mathcal{A} = a_0 \xi L_0^2 r_0/k$ corresponds to the baseline friction. 
Lastly, $\mathcal{D} = D/(L_0^2 r_0)$ is the diffusion of closed bonds.

For the remainder of this article, we use a dimensionless polymerization speed of $v_{\rm{p}}=0.1$, which corresponds to $0.1 \ \mu$m/s as typical order of magnitude experimentally observed for various cell types \cite{barnhart_adhesion-dependent_2011, schreiber_adhesionvelocity_2021}. 
As discussed above, we use $D = 0.1 \ \mu \rm{m}^2$/s leading to a dimensionless diffusion coefficient of $\mathcal{D} = 0.01$.
This and the estimates of the other parameters are available in \ref{sec:Parameter estimates} and imply $\mathcal{L} = 1.0$, $\sigma_{\rm{act}} = 0.1 $ and $\mathcal{F} = 0.16$.
The baseline friction $\mathcal{A}$ is an effective parameter and cannot be directly inferred from experiments. It serves a regularizing function and, as we will show in the following section, should be sufficiently small. In the following we will then use $\mathcal{A} = 0.1$. The binding parameter $\mathcal{R}$ acts as main continuation parameter and is not fixed to a specific value. The expected value is of the order of one.

\subsection{Adhesion-based mechanism for polarization} \label{sec: polarization mechanism}

Without polymerization a homogeneous stress state, determined by the active stress 
$\sigma(u) = \sigma_{\rm{act}}$, satisfies equations Eq.~\ref{eq:nondim BVP stress} and Eq.~\ref{eq:nondim BVP stress BCs} independently of the adhesion density. As a result, there is no actin flow, leading to $\dot{l}_\pm = 0$ and $\tilde{v} = 0$. The steady state of the adhesion is implicitly described by the equation
\begin{equation} \label{eq:only adhesion}
    0 = \mathcal{R} - {\rm{exp}} \left(\frac{1}{\mathcal{F}}\frac{\sigma_{\rm{act}}}{\mathcal{A} + a(u)} \right) a(u) + \frac{\mathcal{D}}{L^2} \partial_u^2 a(u),
\end{equation}
where the BCs simplify to $\partial_u a(u_\pm) = 0$. First, we analyse uniform solutions, $\partial_u a = 0$, such that the diffusion term vanishes. Eq.~\ref{eq:only adhesion} then characterizes the local balance of binding and unbinding. 

\begin{figure}
    \centering
    \includegraphics[width=15cm,keepaspectratio]{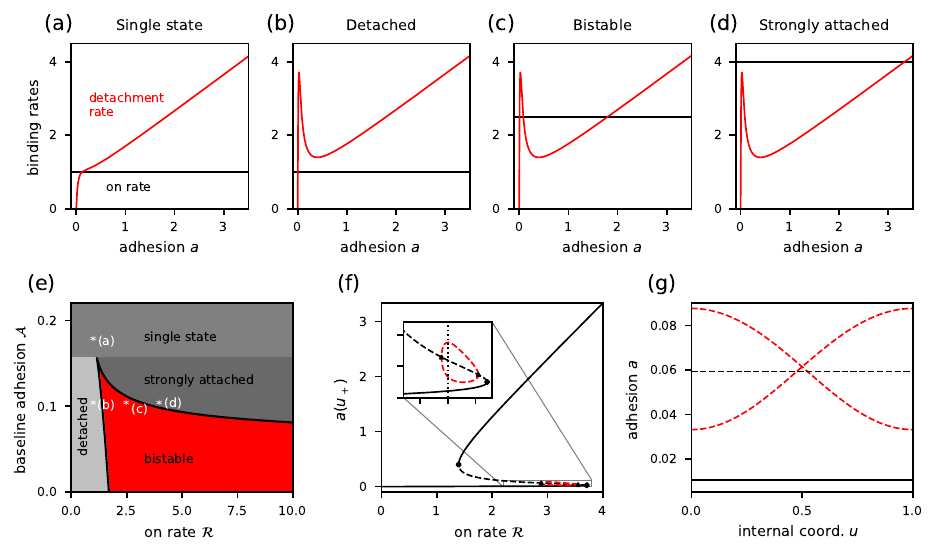}
    \caption[]{Adhesion-based polarization mechanism without polymerization and for constant stress. (a)-(d) Four qualitatively different scenarios for the local steady state adhesion density depending on the baseline friction $\mathcal{A}$, and the on-rate $\mathcal{R}$ (shown as the black horizontal line). (e) The corresponding phase diagram exhibits four regions, where the thick black line indicates the loci of saddle-node bifurcations. 
    (f) Steady state bifurcation diagram including inhomogeneous solutions. The adhesion density at the leading edge $a(u_+)$ is shown as a function of the on-rate $\mathcal{R}$, obtained by numerical continuation. Black lines belong to homogeneous states, red ones are associated with polarized states. Solid/dashed lines indicate stable/unstable states. Diamonds represents saddle-node and triangles pitchfork bifurcations. (g) Corresponding adhesion profiles at $\mathcal{R}=3.0$ in (f). Solid/dashed states are stable/unstable.}
    \label{fig:Polarization mechanism}
\end{figure}

Fig.~\ref{fig:Polarization mechanism}(a)-(d) illustrates the four qualitatively distinct cases which arise in this situation when varying $\mathcal{R}$
(black horizontal lines) and $\mathcal{A}$, i.e.~binding and unbinding. 
In (a), the unbinding rate increases monotonically with $a$ and the only feasible steady state occurs at the intersection of the two curves at low adhesion. 
Since the baseline adhesion $\mathcal{A}$ is large, load sharing is suppressed for small $a$. Reducing baseline adhesion $\mathcal{A}$ below a critical value in Fig.~\ref{fig:Polarization mechanism}(b), 
the detachment rate becomes non-monotonous in $a$ due to load sharing. Nonetheless, with the chosen on-rate $\mathcal{R} = 1.0$, 
only one state is accessible. 
We term this low adhesion state "detached". As the on-rate increases in Fig.~\ref{fig:Polarization mechanism}(c), we enter a regime of multiple solutions through a saddle-node bifurcation. 
Now the system displays bistability between the detached and a strongly attached state. 
With an even larger on-rate in Fig.~\ref{fig:Polarization mechanism}(d), only the strongly attached state persists. 
In Fig.~\ref{fig:Polarization mechanism}(e), we present
the full phase diagram. The thick black lines indicate the loci of saddle-node bifurcations, where two solutions emerge, when transitioning from (b) to (c), and, subsequently, annihilate each other from (c) to (d). This analysis
is similar to the bifurcation analysis of the
standard model for adhesion contacts with mechanosensitive
dissociation rates \cite{bell_models_1978,erdmann_stability_2004,sens_stickslip_2020}, but differs in the existence of a baseline friction.

To investigate the behaviour of steady states in the full, spatially resolved model, we performed numerical continuation of the homogeneous state \cite{Doedel_2007_AUTO07p}.
We consider the adhesion density at the right cell edge $a(u_+)$ in order to capture spatially inhomogeneous and symmetry-broken solutions of equation Eq.~\ref{eq:only adhesion}. 
In the bifurcation diagram in Fig.~\ref{fig:Polarization mechanism}(f),
we plot it as a function of the on-rate $\mathcal{R}$. The black curves represent the discussed homogeneous steady states, where stable (unstable) states are indicated by solid (dashed) lines. 
The existence of this bistable regime forms the basis of the polarization mechanism presented here: different regions of the cell might be locally in balance either in the detached or attached state. 
The diffusion term then introduces a spatial coupling and connects these different regions to form smooth adhesion profiles. Because diffusion generally suppresses spatial inhomogeneities, 
it must be sufficiently slow for such solutions to exist. For the standard value $\mathcal{D}=0.01$ used here, two inhomogeneous branches (red) bifurcate from the unstable homogeneous branch, where the upper (lower) branch is associated with more adhesion at the right (left) edge of the cell. 
The further the system deviates from the homogeneous state, the stronger the polarization becomes. 
Corresponding adhesion profiles at $\mathcal{R}=3.0$ are shown in Fig.~\ref{fig:Polarization mechanism}(g), where the adhesion density at the one cell edge exceeds twice that at the other edge.

The emergence of pitchfork bifurcations and the corresponding solutions' symmetry with respect to the cell's midpoint, reflects the left-right symmetry inherent in the underlying equations. 
This symmetry is subsequently spontaneously broken, with stronger polarization as diffusion slows down. 
As it is typical for diffusive systems, higher order modes with additional peaks emerge when diffusion is reduced further. 
However, in the context of motility, only the first mode, which is depicted here, is relevant.

So far, our analysis has focused only on adhesion, neglecting its coupling to cell mechanics. 
Most importantly, the analysis from Fig.~\ref{fig:Polarization mechanism} indicates that
adhesion might lead to rich dynamics. However, if not complemented by other processes,
the only stable solution is a sessile state without persistent SSB.
Linear stability analysis (cf.~\ref{sec:LSA without polymerization}) shows that without a finite value of the polymerization velocity $v_{\rm{p}}$, 
the adhesion density is a second-order effect around the homogeneous stress base state $\sigma (u) = \sigma_{\rm{act}}$ with $L = 1 - \sigma_{\rm{act}}$. The stress and length relax toward this state independently of the adhesion perturbation, at least up to first order perturbations. In finite volume simulations \cite{Guyer_2009_CompSciEng_FiPy}, we observed a relaxation even for initial conditions far away from this state, in agreement to previous results without the additional adhesion field \cite{drozdowski_optogenetic_2021}. Symmetry broken initial conditions result in a transient directed movement, but the cell eventually comes to a rest because all polarized states are unstable (cf. Fig.~\ref{fig:Polarization mechanism}(g)).
These results indicate that the uniform, non-motile state is globally stable and
the stress must be forced out of its homogeneous state by other means.
We now show that stable SSB and persistent motility can be obtained
by combining the binding dynamics analysed above
with symmetric edge polymerization.

\subsection{Motile solutions in presence of polymerisation}

\begin{figure}
    \centering
    \includegraphics[width=15cm,keepaspectratio]{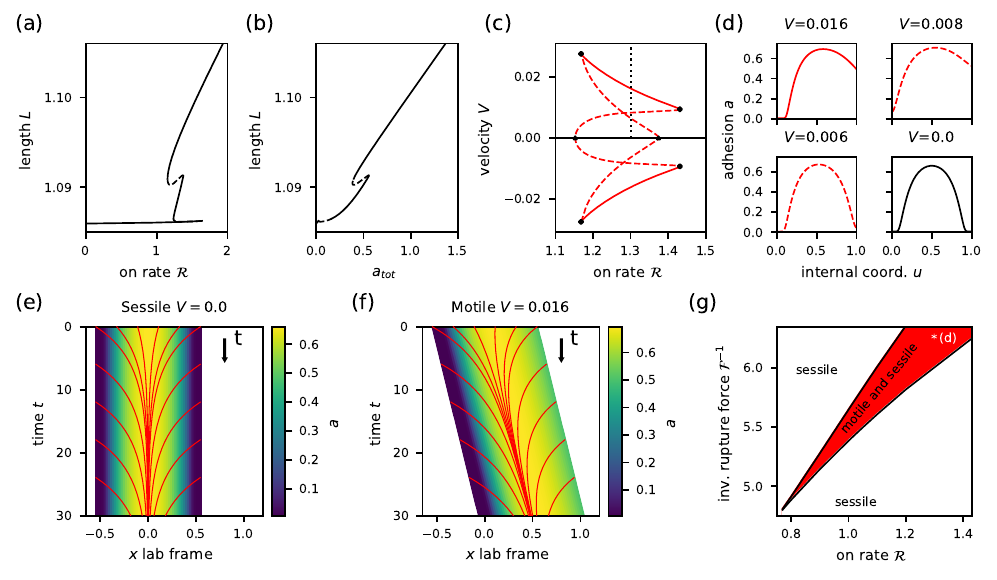}
    \caption[]{Polymerization stabilizes migration in the full model. (a) Steady state cell length $L$ of sessile states as function of the on-rate $\mathcal{R}$ demonstrates multistability. (b) Cell length $L$ as function of the integrated adhesion density $a_{\rm{tot}}$ shows an overall linear relation, but also multistability. For intermediate adhesiveness, three stable sessile states exist: weakly, moderately and strongly attached. (c) Bistability in the cell velocity $V$ between motile states in red and the three sessile states ($V=0$) in black. Diamonds represent saddle-node and triangles pitchfork bifurcations. (d) Corresponding adhesion profiles at $\mathcal{R}=1.3$. (e) Kymograph of the sessile state. Adhesion density is color-coded and retrograde actin flow is indicated by red lines. (f) Kymograph of the stable motile state. (g) Phase diagram in on-rate $\mathcal{R}$ ("binding") and inverse rupture force $\mathcal{F}^{-1}$ ("unbinding") shows bistable region, in which stable migration occurs. Thick black lines indicate the loci of the saddle-node bifurcations on the motile branches (red) in (c).}
    \label{fig:Motility}
\end{figure}

Now considering the full system, equations Eq.~\ref{eq:nondim BVP stress}-Eq.~\ref{eq:nondim BVP adhesion BCs}, any finite polymerization velocity $v_{\rm{p}} > 0$ induces flow at the boundaries. 
Consequently, $\partial_u \sigma(u) \neq 0$, and a constant stress profile is no longer a valid solution of the problem. 
The adhesion density then becomes pertinent in the stress equation Eq.~\ref{eq:nondim BVP stress}, in particular impacting the steady state solutions. 
In this sense, polymerization disrupts the uniform solution and adhesion can emerge as a first order effect around the former uniform steady state. 
Despite the feedback loop between stress $\sigma$ and adhesion $a$ via the mechanosensitive off-rate, the fundamental structure of steady-state solutions in Fig.~\ref{fig:Polarization mechanism}(f), characterized by the presence of multiple solutions and the appearance of asymmetric branches, remains intact. 
Therefore, we can understand the mechanism underlying motility based on the analysis in the preceding section.

In Fig.~\ref{fig:Motility}(a), the cell length $L$ is depicted as a function of the on-rate $\mathcal{R}$, now with a finite
polymerization velocity ($v_{\rm{p}}=0.1$). We observe a similar structural pattern as in Fig.~\ref{fig:Polarization mechanism}(f), where multiple solutions are present within an intermediate range of $\mathcal{R}$. 
Fig.~\ref{fig:Motility}(b) illustrates the relation between $L$ and the total adhesion density $a_{\rm{tot}} = \int_0^1 a(u) \rm{d}u$. 
For very weak adhesion, the length remains almost independent of $a_{\rm{tot}}$ due to the dominance of baseline friction $\mathcal{A} = 0.1$. 
Toward larger adhesion, length increases nearly linearly with adhesion, i.e. our model correctly predicts spreading of adherent cells on typical culture substrates \cite{reinhart-king_dynamics_2005}. 
It is worth noting that even for $a_{\rm{tot}}=0$, the steady-state length has increased compared to the case without polymerization. 
This occurs because both the baseline friction $\mathcal{A}$ and the intrinsic viscosity of the actin network, represented by $\mathcal{L}$, slow down retrograde flow.

Polarized branches are now associated with directed migration. In Fig.~\ref{fig:Motility}(c), 
we illustrate the steady-state velocity $V$ as a function of the on-rate, with motile branches highlighted in red (for reasons of clarity, motile branches are not shown in panel (a) and (b)). In this representation, all symmetric branches collapse to a single line $V=0$. 
Compared to Fig.~\ref{fig:Polarization mechanism}(f), two saddle-node bifurcations occur on each polarized branch, enclosing a stable region. 
The corresponding adhesion profiles at $\mathcal{R}=1.3$ are depicted in Fig.~\ref{fig:Motility}(d), where only one of the three stable sessile solutions is shown (bottom right in black). 
Similarly to before, both a detached and a strongly attached state persist as well. Fig.~\ref{fig:Motility}(e) and (f) show kymographs of the sessile and motile state, respectively, 
obtained by finite volume simulations. 
In both cases, the adhesion profile stays constant over time, confirming that both of them represent steady states of our system. Finally
Fig.~\ref{fig:Motility}(g) shows the full phase diagram, demonstrating that motile solutions are possible if
adhesion is complemented by polymerization.

\subsection{Comparision with experimental observations}

Our theoretical results are in good qualitative agreement with experimental observations. We first note that our system exhibits true bistability between a motile and multiple sessile states, as observed for different cell types migrating on both 2D substrates \cite{verkhovsky_self-polarization_1999, barnhart_adhesion-dependent_2011} and on 1D tracks under lateral confinement \cite{hennig_stick-slip_2020, amiri_multistability_2023}. 
The border between the sessile and bistable regime in the phase space diagram in Fig.~\ref{fig:Motility}(g) is given by the loci of the saddle-node bifurcations on the motile branches. 
As explained before, a minimal inverse rupture force $\mathcal{F}^{-1}$ is required to achieve polarization. 
Because the bistable regime extends along the diagonal in Fig.~\ref{fig:Motility}(g), we conclude
that binding and unbinding have to be in balance to enable motility. Away from this region, only
sessile solutions exist. Below we will investigate
switching between these states by applying external perturbations.

We next note that in the  stable motile states, adhesion is strongly
polarized, with strong adhesion at the front and low adhesion at the back, as observed
in experiments \cite{fournier_force_2010, barnhart_balance_2015, schreiber_adhesionvelocity_2021}. 
This asymmetry in adhesion distribution leads to the differential rates of retrograde flow between the front and back regions of the cell, facilitating effective forward propulsion through polymerization-driven membrane protrusion. 
As new adhesive bonds need time to form in the flow, adhesion density peaks near the cell center. It slightly diminishes toward the leading edge as observed, for instance, in MDA-MB-231 cells \cite{schreiber_adhesionvelocity_2021}. 
Actomyosin contraction and membrane tension drive retrograde flow, surpassing the speed of polymerization at the back of the cell, 
allowing for efficient rear retraction with minimal adhesive resistance. 
Upon comparison with unstable states, it becomes apparent that both an overall adhesion polarity and the absence of adhesion at the rear are crucial for sustaining efficient cellular migration.

We finally note that in mesenchymal motility, a biphasic adhesion–velocity relation has been measured across various cell types, 
representing a universal principle in cell migration, where a maximum migration velocity occurs at intermediate ligand concentrations on the substrate \cite{palecek_integrin-ligand_1997, barnhart_adhesion-dependent_2011, schreiber_adhesionvelocity_2021}. 
In our model, motility is only feasible in a regime of intermediate on-rate $\mathcal{R}$, recapitulating the property of an optimal adhesiveness: 
if adhesion is too low (detached state), active processes solely generate retrograde flow, but fail to transmit their forces to the substrate. 
Conversely, excessive adhesion prevents the detachment of adhesive bonds at the rear, resulting in a stalled cell. 
Accordingly, increasing (decreasing) the polymerization speed, as one of the two force-generating processes, shifts the motile regime to larger (smaller) values of the on-rate. 
Even minor adjustments in $v_{\rm{p}}$ can significantly impact the range of motility in $\mathcal{R}$ due to the exponential coupling between stress and detachment rate.

\section{Effect of external perturbations in adhesion} \label{sec:Induced failure of adhesion bonds}

To demonstrate the bistable nature of our model and the role of adhesion in motility initiation, we investigate switching from a sessile to a migrating state by applying an asymmetric perturbation to the adhesion density. 
This perturbation is implemented in a finite volume simulation by setting the adhesion density to zero over a certain fraction of the cell length  at a certain point in time ($t=0$), corresponding to a complete loss of adhesion to the substrate, and observing the temporal evolution. 
From the experimental perspective, this means to disrupt adhesion without destroying the actin network itself or impair the adhesiveness of the substrate, such that the on-rate $\mathcal{R}$ in our model remains the same.

\begin{figure}
    \centering
    \includegraphics[width=15cm,keepaspectratio]{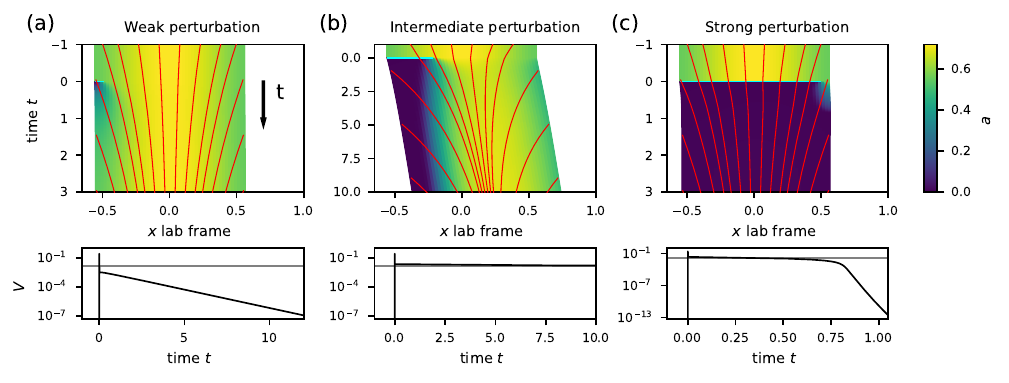}
    \caption[]{Motility initiation via external perturbation: starting in the strongly attached state, adhesion is destroyed over the left (a) 5 $\%$, (b) 30 $\%$ and (c) 95 $\%$ of the cell length at $t=0$. The upper panels show the corresponding kymographs in the lab frame $x$ with color-coded adhesion density. The actin flow is indicated as red lines, the perturbation as horizontal bars in cyan. Lower panels show the cell velocity as function of time, where the grey line represents the stable migratory steady state.}
    \label{fig:Induced failure}
\end{figure}

In Fig.~\ref{fig:Induced failure}(a) adhesion is "erased" on a small interval of 5~$\%$ of the initial cell length at the left edge. Because friction is drastically reduced in this area, retrograde flow increases and causes a transient, small retraction of this non-adhesive edge (upper panel). 
This is accompanied by an immediate positive spike in the cell velocity (lower panel), corresponding to the onset of movement to the right, i.e.~to the side of stronger adhesion. 
However, adhesion manages to quickly recover because the overall adhesion density is only slightly affected. 
The recovery is only possible due to the spatial coupling and the diffusion of adhesion from strongly to weakly attached regions. Simultaneously, the velocity decreases exponentially to zero (lower panel), exactly as one would expect from linear stability analysis in the proximity of a stable state. 
The velocity actually exceeds the steady migration velocity immediately after the perturbation, but cell length and adhesion profile are too far from the migratory state.

To escape the basin of attraction of the sessile state, the applied perturbation has to be stronger. 
By disrupting adhesion over 30~$\%$ of the cell length in the prospective rear, as shown in Fig.~\ref{fig:Induced failure}(b), a non-local effect can be seen in the actin flow, which exhibits a kink toward the right even far from the perturbed region. 
Thus, an overall movement to the right is initiated. 
Interestingly, the adhesion density on the non-perturbed edge is also reduced because new bonds have to form at the very leading edge during migration, as explained in the previous section. 
Both effects are facilitated by the non-local coupling of the cell edges through the membrane tension, represented in the symmetric stress BCs Eq.~\ref{eq:nondim BVP stress BCs}. 
The initial spike in velocity is actually of the same magnitude as for the weak perturbation. 
However, instead of subsequently decreasing toward zero, it approaches the steady migrating velocity (lower panel). 
The adhesion recovers to a small degree closer to the cell middle and then develops into the stable state (cf.~Fig.~\ref{fig:Motility}(d), top left) with a non-adhesive rear and an overall polarization.

A very strong perturbation, where adhesion is erased over 95~$\%$ of the cell length, as investigated in Fig.~\ref{fig:Induced failure}(c), leads again to a initial velocity spike of the same order as for the two previous cases. The cell
enters a transient state, during which it migrates with a similar velocity as in (b), but the adhesion profile is quite different.
Eventually, the remaining adhesion at the leading tip cannot withstand the active forces and diminishes rapidly, in the shown example around $t=1$. 
Therefore, the velocity decays exponentially and the movement stops (lower panel). In contrast to panel (a), the cells ends up in the detached state. It should be noted that the bulk properties are not immediately reflected in the current cell length and velocity right after 
the perturbation, but they are highly relevant in determining the long-term behaviour and final state. This prediction is only possible in a spatially resolved model and highlights the benefits of our continuum approach compared to previously proposed point-like descriptions.

In conclusion, a sufficiently strong polarization has to be established to switch to the motile state without reducing the overall adhesion density too much. Then, the cell edge losing adhesion becomes the prospective trailing rear. 
In practise, cells are subject to strong thermal noise \cite{hennig_stick-slip_2020,amiri_multistability_2023}, which can cause the switching between 
states, here shown in a purely deterministic theory. 
In principle, optogenetics could be used for such manipulations as well and have been recently applied to regulate talin as a mechanical linker in cell-matrix adhesions \cite{sadhanasatish_molecular_2023}. 
The instantaneous disruption of adhesion can easily be extended to a time-integrated signal resembling the effect of optogenetics within our framework.

Our observations demonstrate the pivotal role of adhesion in establishing global polarity and initiating motility without previous polarization of the actomyosin system itself. 
Furthermore, it supports the hypothesis that the cell rear can be defined by sudden detachment \cite{hennig_stick-slip_2020}. 
Because in this purely mechanical model without a direct or indirect positive correlation between polymerization and adhesion assembly, edge polymerization alone would not be able to create a stable leading edge. 
Even though an asymmetric increase in polymerization would indeed create a transient protrusion and migration in the corresponding direction, the created stresses accelerate adhesion detachment at this side. 
As soon as polymerization goes back to its symmetric configuration, the movement would stop or even revert. 
To compensate this effect one could consider catch-bond behaviour for small stresses, where the off-rate decreases under small load, or directly couple adhesion and polymerization. Such an interaction was experimentally observed \cite{vicente-manzanares_integrins_2009} and could be mediated e.g. by the Arp2/3 complex, facilitating both actin polymerization and adhesion formation \cite{goley_arp23_2006, sever_actin_2018}. 

\section{Effect of patterning adhesion}
\label{sec:patterned_adhesion}

Finally, we analyse the behaviour of cells in a structured environment, including adhesive steps 
\cite{barnhart_adhesion-dependent_2011,schreiber_adhesionvelocity_2021} and continuous gradients of adhesiveness \cite{ricoult_substrate-bound_2015}. 
We introduce a spatial dependence of the on-rate and assume that this adhesive cue is substrate-bound and stays constant over time, i.e. it is neither consumed or produced by the cell itself. 
The guiding by such an adhesive cue is called haptotaxis.

\subsection{Haptotaxis on adhesive gradients}

\begin{figure}
    \centering
    \includegraphics[width=15cm,keepaspectratio]{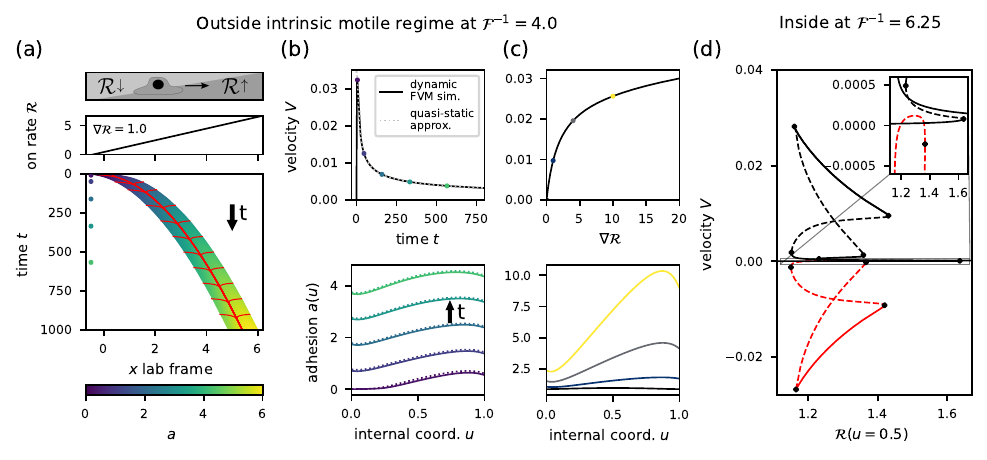}
    \caption[]{Haptotaxis on continuous gradients. (a)-(c) Outside the intrinsic motile regime. (a) The upper panel shows a sketch of a cell on a constant gradient of the on-rate, which is depicted in the middle panel. The lower panel displays the kymograph with color-coded adhesion density and actin flow in red. (b) Corresponding cell velocity (top) and adhesion profiles at different time points (bottom). The quasi-static approximation is obtained by numerical continuation. (c) Quasi-static approximation of the velocity as a function of the gradient (top) and corresponding adhesion profiles (bottom) for fixed on-rate at the trailing edge of $\mathcal{R}(u_-)=1.3$. (d) Bifurcation diagram of the steady state velocity $V$ within the intrinsic motile regime for a small external gradient in the on-rate of $\nabla \mathcal{R}=0.01$. The x-axis shows the values of the on-rate at the cell middle $u = 0.5$. This perturbation results in imperfect bifurcations (inset) and disconnected branches (black and red).}
    \label{fig:External gradient}
\end{figure}

To demonstrate the basic principles of haptotaxis in our model, we first 
analyse the behaviour on constant gradients within and outside the intrinsic motile regime. 
In Fig.~\ref{fig:External gradient}(a)-(c), we choose a characteristic rupture force of $\mathcal{F}^{-1}=4.0$ such that, regardless of the on-rate, only one sessile state exists (cf. Fig.~\ref{fig:Motility}(g)). 
Nevertheless, the cell is able to follow the gradient in the on-rate uphill (here $\nabla \mathcal{R}=1.0$), because the heterogeneous environment creates polarization within the initially symmetric cell.
The lower part of panel (b) shows the adhesion profiles over time. Since the gradient is constant, the polarity and shape of these profiles are almost identical, neglecting an overall adhesion density increase. 
The cell length increases, while the migration speed (upper panel in (b)) deteriorates. An increased adhesion must be overcome at the cell rear, impeding retraction in the rear. 
Additionally, larger stress gradients facilitate retrograde flow at both edges. Thus, we observe the same trend as in the motile regime, 
where the largest velocities occur for the smallest possible on-rate (cf. Fig.~\ref{fig:Motility}(c)).

The dynamical behaviour, obtained with finite volume simulations, can be approximated in a "quasi-static" manner (dashed lines in Fig.~\ref{fig:External gradient}(b)). 
Numerical continuation allows us to predict the steady state corresponding to the current environment, namely the gradient and the absolute on-rate e.g. at the cell middle. 
The cell will never reach this state in a dynamical situation because the absolute on-rate changes permanently during migration.
However, since both the velocity curves and adhesion profiles agree very nicely after an initial phase, we can turn the setup in the quasi-static approximation around and predict the velocity and adhesion profiles as a function of the gradient given the absolute on-rate value. 
The polarity increases with the gradient as expected (cf. lower panel in Fig.~\ref{fig:External gradient}(c)). However, we observe a nonlinear velocity-gradient relation, where the increase in velocity decreases for larger gradients. 
The overall growth in adhesion density limits the increase in velocity. 
In particular, adhesion gets significantly stronger at the trailing edge, even though the on-rate is kept constant there. This is mainly caused by diffusion, which becomes more relevant for larger adhesion values, and to a smaller amount by the effective advection.

Applying a small gradient within the intrinsic motile regime for $\mathcal{F}^{-1}=6.25$ perturbs the bifurcation diagram, as illustrated in Fig.~\ref{fig:Motility}(d).
Any small but finite gradient breaks the underlying left-right symmetry and, therefore, converts the former pitchfork to imperfect bifurcations, 
such that the positive and negative velocity branches become disconnected (cf. inset in Fig.~\ref{fig:External gradient}(d) for $\nabla \mathcal{R}=0.01$). 
This is accompanied by an overall shift in velocity, in this case upward since the gradient promotes migration to the right. 
However, the cell can still migrate against such a small gradient. Above a critical value, which is around $\nabla \mathcal{R} \approx 0.1$, this is no longer possible. 
Interestingly, the maximum possible migration speeds of both intrinsic and externally driven migration as well as their combination are always of the same order of around $0.03$. 
This highlights the fact that the velocity is determined by the forces generated within the cell, which are in turn controlled by the active processes, 
which we kept constant here. Thus, even though external cues can steer migration, it is powered by the cell itself.

\subsection{Motility initiation on lines with adhesive steps}

\begin{figure}
    \centering
    \includegraphics[width=15cm,keepaspectratio]{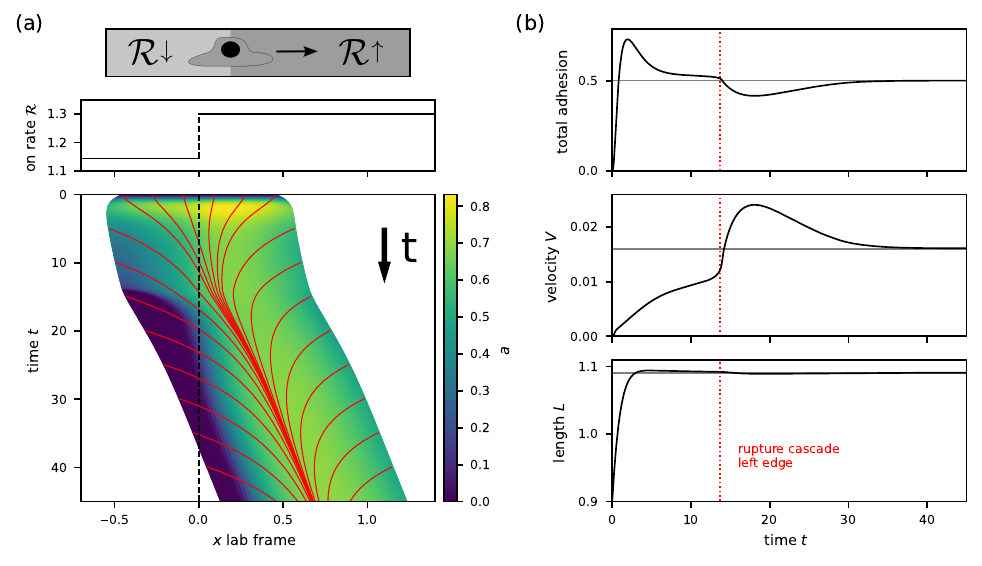}
    \caption[]{Polarization and motility initiation on an adhesive step. (a) Sketch of the protocol (top), spatially dependent on-rate (middle) and kymograph with color-coded adhesion density and actin flow indicated by red lines (bottom). The cell is initialized without any adhesion and a length of $L=0.9$. (b) Corresponding integrated adhesion density (top), cell velocity $V$ (middle) and cell length $L$ (bottom) as functions of time. Grey lines represent the stable migratory steady state.}
    \label{fig:Motility initiation pattern}
\end{figure}

After having established the basic haptotactic behaviour of cells preferring stronger adhesiveness, we now consider an initially completely non-adherent cell on top of an adhesive step. 
This is schematically illustrated in the top panel of Fig.~\ref{fig:Motility initiation pattern}(a) with the corresponding on-rate $\mathcal{R}$ below. 
For the larger on-rate on the right side of the step, the system is within the multistable regime with a stably migrating steady state. 
On the left side, the system is slightly below the critical value, such that the only available steady state is the detached state. 
In the beginning at $t=0$, adhesive bonds form almost homogeneously over the complete cell length and spreading occurs symmetrically to both sides due to polymerization, which in turn drives retrograde actin flow. 
When the cell reaches its maximum length (cf.~bottom panel in (b)) and therefore the boundary stress peaks, the adhesion density stagnates and subsequently decreases. 
However, while on the right side a stable density is achieved, the left half of the cell loses more and more traction. 
This is accompanied by a slight shift in actin flow to the right, until a critical density on the left edge is reached and a rupture cascade causes the sudden loss of adhesion at the rear, in the shown example at around $t=14$, resulting in a steep increase of velocity. 
A global polarization has now been established and the cell migrates to the right, as expected.

When the cell gradually enters into a region of higher adhesion density, the non-adhesive rear shrinks in size and overall adhesion rises again. 
However, even after the cell has completely crossed the step, polarity remains intact in a now homogeneous environment. 
Therefore, a sufficiently steep step in the correct regime of adhesiveness provides another initiation mechanism to switch into the migratory state. 
Key to this permanent polarization is that on the less adhesive side only the detached state is available, while the right part is within the motile regime (cf. Fig.~\ref{fig:Motility}(c)). 
Based on our findings in Sec.~\ref{sec:Induced failure of adhesion bonds} on motility initiation via an external perturbation, we conclude that a sufficiently large fraction has to be exposed to the less adhesive region to form the prospective non-adhesive rear. 
If this was not the case, the cell would still show haptotactic behaviour by sensing and entering the more adhesive region, but would stop right after the step.

\subsection{Motility arrest and direction reversal}

After studying motility initiation, we now turn to the opposite process, when a migrating cell approaches a step 
downwards in adhesiveness. 
In principle, three distinct scenarios are known from experiments \cite{schreiber_adhesionvelocity_2021, amiri_multistability_2023}: the cell continues to migrate in the same direction, the movement is stopped or the direction is reversed under repolarization of the cell. 

\begin{figure}
    \centering
    \includegraphics[width=15cm,keepaspectratio]{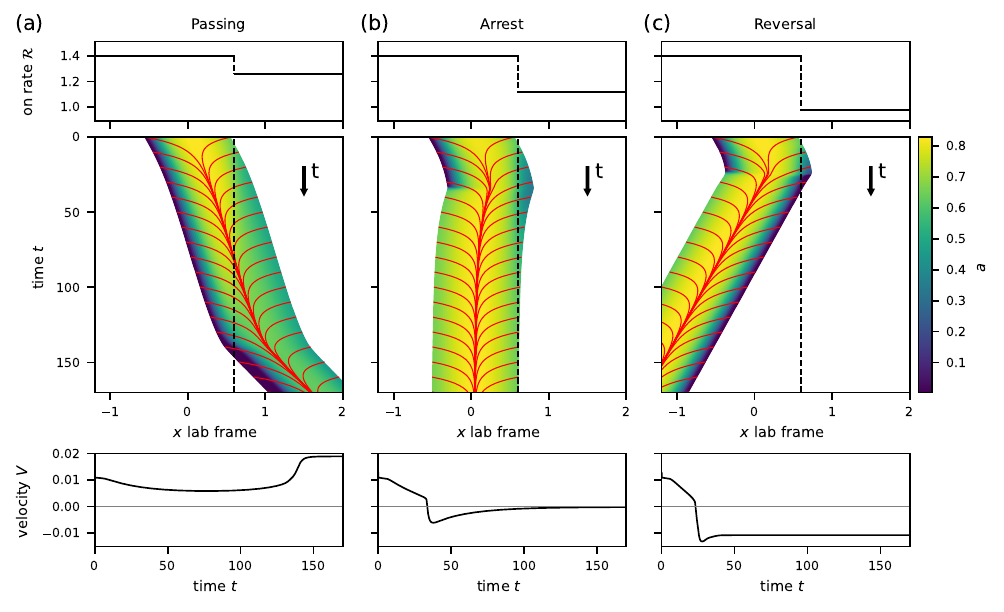}
    \caption[]{Motility arrest and reversal at an adhesive step. A stably migrating cell approaches a step in the on-rate (top) from $\mathcal{R}=1.4$ down to (a) 90 $\%$, (b) 80 $\%$ and (c) 70 $\%$. Kymographs show color-coded adhesion density and actin flow indicated by red lines. The cell will pass small steps, migration stops for intermediate steps and it is reverted under repolarization of the adhesion density for large steps. The bottom row shows the corresponding cell velocity $V$ over time. The sessile state with $V=0$ is indicated in grey.}
    \label{fig:Pattern reversal}
\end{figure}

All three cases occur in our model, as demonstrated in Fig.~\ref{fig:Pattern reversal}, depending on the initial migrating state and on the step size. 
In Fig.~\ref{fig:Pattern reversal}(a), the on-rate is reduced by 10~$\%$ across the step. 
While the cell passes the step, the adhesion density is gradually reduced from front to back leading to a transient weakening of the polarization and therefore a reduction of the velocity. 
Once the cell has completed the step, overall adhesion is obviously less compared to the initial state, but the velocity has increased, in agreement with the bifurcation diagram in Fig.~\ref{fig:Motility}(c), where motile states are faster for smaller on-rates.

In Fig.~\ref{fig:Pattern reversal}(b) we observe a motility arrest for a step size of 20~$\%$. 
The on-rate of $\mathcal{R}=1.12$ after the step is not within the motile regime anymore, such that complete passing is not achievable. The leading edge transiently crosses into the less adhesive region, similar to the "passing" case in panel (a). 
However, adhesion at the cell front is significantly more reduced, facilitating retrograde flow. 
Thus, protrusion speed is slowed down, even before the polarization is lost, concomitantly with a shortening of the cell length. 
This in turn reduces the stress at the cell edges, until a critical point around $t=30$. 
Then, the on-rate excels the off-rate at the rear, leading to a rapid growth in adhesion and an arrest of the directed movement. 
In agreement with our observation of haptotaxis in the previous section, the unpolarized cell prefers the highly adhesive region and slowly returns to the left side of the step. 
The observed fast adaptation demonstrates the local cooperativity of adhesive bonds. 
In contrast to an advective mechanism based on intracellular transport, this allows a cell to sense its environment very fast and efficiently.

For complete repolarization and reversal of direction, the former front of the cell must switch to the detached state. 
Increasing the step size further to 30~$\%$ leads to the expected effect, as shown in Fig.~\ref{fig:Pattern reversal}(c): 
subsequently to the formation of strong adhesion at the former rear, the cell length and the stress at the edges start to grow again, 
causing a rupture cascade at the right edge and a reversal of actin flow. 
The repolarized cell migrates in the opposite direction and polarity remains intact after crossing back into a homogeneous environment. 
Larger steps accelerate the repolarization process but result in the same qualitative behaviour of reversal. 
This behaviour cannot be predicted solely by the known steady states, since the dynamical length changes play a crucial role, demonstrated by the qualitative difference between (b) and (c), where in both cases the only accessible steady state in the less-adhesive region is the sessile, detached state. 
However, only in (c) we observe full repolarization. 
The intricate interplay of length, velocity, and cell fraction beyond the step boundary, with the effected bond rupture behavior, necessitates a one-dimensional representation of the cell.
This constitutes a strength of our spatially resolved one-dimensional approach. 
For every pair of left and right on-rates a dynamical simulation can be conducted to obtain predictions.

\section{Discussion}

In this work, we have shown how active gel theory can be extended by a dynamic adhesion field in a 1D setting. 
Similarly to the molecular clutch model, actin flow is slowed down by adhesive contacts to the substrate; but differently from standard molecular clutch
models, the direction of flow is not assumed, but predicted by our theory. 
Our model demonstrates that the interplay
between local load sharing between mechanosensitive bonds and initially symmetric polymerization-driven retrograde flow
leads to spontaneous symmetry breaking. Thus our theory identifies a novel polarization and motility mechanism for 1D cell migration. 
Because the model gives clear predictions on actin flow and adhesion dynamics, 
it can be compared directly to experiments. In fact, the predicted ranges of migration speeds and adhesion profiles are in very good agreement with experimental measurements of cells on 1D lines. 
In agreement
with experimental observations, our model exhibits bistability between sessile and motile states. 
We have demonstrated how local perturbations in adhesion can induce switching and have highlighted the potential role of adhesion in motility initiation. 
While the adhesion density close to the boundary determines the current edge movement, bulk properties dictate the long-term behaviour and convergence toward a steady state. This distinction is only possible due to our continuum framework.

By introducing a spatial dependence of the binding rate in our model, we are able to describe heterogeneities and  especially gradients and steps in substrate adhesiveness. Our model qualitatively captures the haptotactic behaviour of cells to prefer more adhesive regions. 
On continuous gradients, 
we have analysed migration within and outside of the intrinsic motile regime. 
In a quasi-static approximation we have access to the full bifurcation diagram. 
Predicted velocities and adhesion profiles agree very well with full dynamical simulations. 
For discontinuous steps in the adhesiveness, we were able to qualitatively reproduce experimentally observed directional reversal and repolarization. 
While the cooperativity of bonds is very sensitive to local environmental changes, the non-local elastic boundary condition provides a fast communication between both cell ends. 
Together, these two mechanism allow for a fast decision making on structured patterns. 
In addition to patterned substrates as external stimulus, direct manipulation of adhesion bonds by means of optogenetics \cite{sadhanasatish_molecular_2023} presents another exciting possibility, which could be included in our model, to better understand the dynamic organization of adhesion.

Since we have assumed an active stress that is constant in space and time, actomyosin contraction only acts in the background and does not drive
SSB by itself. However, adding a dynamic myosin concentration field, as done e.g.\ in reference \cite{lo_mechanosensitive_2024}, could lead to more complex behaviour, such as oscillatory stick-slip migration \cite{hennig_stick-slip_2020, amiri_multistability_2023}. Furthermore, it would become possible to compare the timescales needed to establish adhesion and myosin polarity. This could greatly enhance our understanding of SSB in real cells and how the different components of the cytoskeleton interact. On the other hand, one could incorporate more details into the binding dynamics, e.g.~by considering catch-bond behaviour \cite{marshall_direct_2003, sens_stickslip_2020} or mechanosensitive adhesion bond recruitment from reservoirs \cite{braeutigam2022generic}, which both might play a role in the interplay of actin flow and maturation of focal adhesions \cite{vicente-manzanares_integrins_2009}. Finally, it would be interesting to
extend the 1D-theory presented here to 2D, for example to
describe the movement of keratocytes and lamellipodial fragments \cite{PhysRevLett.110.078102,reeves_rotating_2018,PhysRevE.105.024403}. 

\section*{Acknowledgements}
This work is funded by the Deutsche Forschungsgemeinschaft (DFG, German Research Foundation) under Germany's Excellence Strategy EXC 2181/1 - 390900948 (the Heidelberg STRUCTURES Excellence Cluster) and by the Max Planck School Matter to Life supported by the German Federal Ministry of Education and Research (BMBF) in collaboration with the Max Planck Society.

\appendix
\section{Derivation of boundary conditions for the adhesion density} \label{sec:Conservation adhesion sites}
We consider two species: bound $a_{\rm{b}} (x,t)$ and unbound $a_{\rm{u}} (x,t)$ adhesion sites. Both densities change due to (un)binding, diffusion and, in addition, unbound sites are advected with the actin  gel's velocity $v$. The dynamic equations then read
\begin{eqnarray} \label{eq:full reaction-diff-system bound} 
\partial_t a_{\rm{b}} &= +\tilde{r}_{\rm{on}} a_{\rm{u}} - r_{\rm{off}} a_{\rm{b}} + D_{\rm{b}} \partial_x^2 a_{\rm{b}}, \\
\label{eq:full reaction-diff-system unbound}
\partial_t a_{\rm{u}} + \partial_x (v a_{\rm{u}}) &= -\tilde{r}_{\rm{on}} a_{\rm{u}} + r_{\rm{off}} a_{\rm{b}} + D_u \partial_x^2 a_{\rm{u}}.
\end{eqnarray}
Without turnover, the total number of adhesion sites $\int_{l_-}^{l_+} a_{\rm{tot}}  dx$, with $a_{\rm{tot}} = a_{\rm{b}} + a_{\rm{u}}$, has to be conserved. 
To take into account the movement of the boundaries, we have to apply the Leibniz integral rule. Then, the conservation condition implies
\begin{eqnarray}
    0 &= \frac{d}{dt} \int_{l_-}^{l_+} a_{\rm{tot}}  dx \\
    &= \dot{l}_+ a_{\rm{tot}} (l_+) - \dot{l}_- a_{\rm{tot}} (l_-) + \int_{l_-}^{l_+} \partial_t a_{\rm{tot}}  dx.
\end{eqnarray}
The local change of the total adhesion density is given by the sum of equations Eq.~\ref{eq:full reaction-diff-system bound} and Eq.~\ref{eq:full reaction-diff-system unbound}. The binding terms cancel each other and we are left with
\begin{eqnarray}
0 = \dot{l}_+ a_{\rm{b}} (l_+) - \dot{l}_- a_{\rm{b}} (l_-) + D_{\rm{b}} \partial_x \left( a_{\rm{b}} (l_+) - a_{\rm{b}} (l_-) \right) \\
\qquad + D_{\rm{u}} \partial_x \left( a_{\rm{u}} (l_+) - a_{\rm{u}} (l_-) \right).\nonumber
\end{eqnarray}
Now, assuming an infinite reservoir of unbound sites, as explained in Sec.~\ref{sec:Minimal model}, the second line of the above equation can be neglected. As a result, we obtain Robin-type BCs for $a_{\rm{b}}$ at each cell edge
\begin{equation}
    \partial_x a_{\rm{b}} (l_\pm) = - \frac{\dot{l}_\pm}{D_{\rm{b}}} a_{\rm{b}} (l_\pm),
\end{equation}
with the ratio of membrane movement and diffusion. By redefining $r_{\rm{on}} = \tilde{k}_{\rm{on}} a_{\rm{u}}$, we obtain Eq.~\ref{eq:Adhesion evolution} from Eq.~\ref{eq:full reaction-diff-system bound}.

\section{Parameter estimates} \label{sec:Parameter estimates}
The following estimates are made for a thin two-dimensional slab of material. The cell is assumed to be homogeneous in the direction orthogonal to its direction of movement, such that our model is effectively a one-dimensional problem. Nevertheless, stresses are still given in units of force per area and bond density is also measured per area. This allows also for easier comparison to experimental measurements.

The cell parameters 
(cf.~Tab.~\ref{tab:physical parameters} top) are taken from either experimental measurements or from similar models, except for the effective friction coefficient, which is derived below. The off-rate without load $r_0$ and the typical rupture force $f_0$ of bonds 
(cf.~Tab.~\ref{tab:physical parameters} bottom) have already been estimated in other models, partly based on experimental measurement. To obtain a reasonable value for the effective friction coefficient $\xi$ in our model, we consider the intermediate friction regime from reference \cite{barnhart_adhesion-dependent_2011}, $\tilde{\xi} = 10^{15}$ Pa s/m$^2$. The typical order of stress in our system is given by the active contractility of $\sigma_{\rm{act}} = 10^3$ Pa. In a motile state, the force per bond should be in the order of the typical rupture force of $f_0$ = 2 pN, such that sufficiently strong binding can occur without stalling the cell. Then, we can estimate the bond density
\begin{equation}
    a = \frac{\sigma_{\rm{act}}}{f_0} = 5 \cdot 10^2 / \mu \rm{m}^2,
\end{equation}
which is the same order of magnitude as experimentally measured \cite{wiseman2004spatial}.
From this value, we obtain the friction coefficient
\begin{equation}
    \xi = \frac{\tilde{\xi}}{a} =  2 \ \rm{ Pa \ s}.
\end{equation}
Under steady state conditions, binding and unbinding should be balanced
\begin{equation}
   r_{\rm{on}} = a \cdot r_{\rm{off}} \approx a \cdot r_0 = 50 /(s \ \mu \rm{m}^2).
\end{equation}
Since we use the on-rate as our continuation parameter, this approximation just provides a rough estimation of the expected order of magnitude.

The diffusion constant $D$ must be smaller than a critical value to allow for polarization, which is the case for the chosen value of $0.1 \ \mu \rm{m}^2$/s, while it is still larger than experimental measurements suggest for integrins in adhesions \cite{wiseman2004spatial}. However, one should consider that our model is 1D, while the measured value corresponds to a 2D diffusion process.
The baseline bond density $a_0$ is an effective parameter, without a direct accessible counterpart in real cells. As discussed in the Sec.~\ref{sec:Model definition}, it stabilizes the system and is chosen sufficiently small to allow for bistability. The chosen value of $a_0 = 50 / \mu \rm{m}^2$ would correspond to 10 $\%$ of the expected dynamical adhesion bond density $a$.

\begin{table}[h!]
    \centering
    \begin{tabular}{|c|c|c|}
       \multicolumn{3}{c}{\textbf{Cell parameters}} \\
       \hline
       Actin polymerization speed $v_{\rm{p}}$ & 0.1 $\mu$m/s & measured in \cite{pollard_cellular_2003, barnhart_adhesion-dependent_2011} \\
       Cell cortex stiffness $k$ & $10^4$ Pa & estimated in \cite{recho_mechanics_2015} \\
       Contractility $\sigma_{\rm{act}}$ & $10^3$ Pa & used in \cite{recho_mechanics_2015, drozdowski_optogenetic_2021} based on \cite{barnhart_adhesion-dependent_2011, oakes_optogenetic_2017} \\
       Bulk viscosity $\eta$ & $10^5$ Pa s & estimated in \cite{barnhart_adhesion-dependent_2011, recho_mechanics_2015} \\
       Cell rest length $L_0$ & 10 $\mu$m & typical order e.g. in \cite{verkhovsky_self-polarization_1999} \\
       Effective friction coefficient $\xi$ & 2 Pa s & estimated based on \cite{barnhart_adhesion-dependent_2011} \\
       \hline
       \multicolumn{3}{c}{\textbf{Bond parameters}} \\
       \hline
       On-rate $r_{\rm{on}}$ & 50 /($s \ \mu \rm{m}^2$) & estimated in this work, \\
       & & similar considerations in \cite{sens_stickslip_2020} \\
       Off-rate without load $r_0$ & 0.1 /s & similar values used in \cite{barnhart_adhesion-dependent_2011, sens_stickslip_2020, ron_one-dimensional_2020}, \\
       & & measured in \cite{lele_investigating_2008} \\
       Typical rupture force $f_0$ & 2 pN & measured in \cite{jiang_two-piconewton_2003, bangasser_determinants_2013}, \\
       & & similar values in \cite{sens_stickslip_2020, ron_one-dimensional_2020} \\
       Baseline bond density $a_0$ & 50 /$\mu \rm{m}^2$ & estimated in this work \\
       Diffusion D & 0.1 $\mu$m$^2$/s & estimated in this work, \\
       & & similar values used in \cite{barnhart_adhesion-dependent_2011, chen_interplay_2023} \\ \hline
    \end{tabular}
    \caption{Physical parameters assumed in our model.}
    \label{tab:physical parameters}
\end{table}

\begin{table}[h!]
    \centering
    \begin{tabular}{|c|c|}
       \multicolumn{2}{c}{\textbf{Cell parameters}} \\
       \hline
       Actin polymerization speed $v_{\rm{p}}$ & 0.1 \\
       Contractility $\sigma_{\rm{act}}$ & 0.1 \\
       Relative viscous time scale $\mathcal{L}$ & 1.0 \\
       \hline
       \multicolumn{2}{c}{\textbf{Bond parameters}} \\
       \hline
       On-rate $\mathcal{R}$ & continuation parameter, expected range 0.1 - 10.0 \\
       Rupture force $\mathcal{F}$ & 0.1 (0.16 used in simulations) \\
       Baseline bond density $\mathcal{A}$ & 0.1 \\
       Diffusion $\mathcal{D}$ & 0.01 \\ \hline
    \end{tabular}
    \caption{Dimensionless parameters derived from the physical parameters given in Tab.~\ref{tab:physical parameters}.}
    \label{tab:dimless parameters}
\end{table}

\section{Linear stability analysis without polymerization} \label{sec:LSA without polymerization}
We consider the uniform steady state with stress $\sigma(u) = \sigma_{\rm{act}}$ and length $L=1-\sigma_{\rm{act}}$, which is always a solution of the equations Eq.~\ref{eq:nondim BVP stress}-Eq.~\ref{eq:nondim BVP flow BCs}, independent of the values of the other parameters and of the adhesion density $a$, which is implicitly given by equation $\mathcal{R} = {\rm{exp}} \left(\frac{1}{\mathcal{F}}\frac{\sigma_{\rm{act}}}{\mathcal{A} + a} \right) a$. 

We now want to analyze the behavior of the stress, when applying small perturbations $\delta \sigma(u,t)$, $\delta a(u,t)$ and $\delta L(t)$. Plugging these perturbations into equation Eq.~\ref{eq:nondim BVP stress} yields
\begin{equation}
    \frac{\mathcal{L}^2}{\left(L + \delta L \right)^2} \partial_u \frac{\partial_u \left(\sigma + \delta \sigma \right)}{\mathcal{A} + a + \delta a} = \delta \sigma.
\end{equation}
Since the considered steady state is uniform in $\sigma$ and $a$, in first order of $\delta$, the expression simplifies to
\begin{equation}
     \frac{\mathcal{L}^2}{L^2} \frac{\partial_u^2  \delta \sigma}{\mathcal{A} + a} = \delta \sigma,
\end{equation}
which is independent of $\delta a$. Therefore, the general solution is given by $\delta \sigma(u,t) = A(t) \cosh \left( \gamma (u - 0.5) \right) + B(t) \sinh \left( \gamma (u - 0.5) \right)$ with the definition $\gamma = L/\mathcal{L} \ \sqrt{(\mathcal{A}+a)}$. The coefficients $A$ and $B$ are determined by the perturbed boundary condition Eq.~\ref{eq:nondim BVP stress BCs}
\begin{equation}
    \delta \sigma (u_\pm,t) = - \delta L(t).
\end{equation}
Then, we can derive
\begin{equation} \label{eq:solution LSA stress}
    \delta \sigma(u,t) = - \frac{\delta L(t)}{\cosh \left( \gamma/2 \right)} \cosh \left( \gamma (u-0.5) \right).
\end{equation}
Plugging the derivative of equation Eq.~\ref{eq:solution LSA stress} into the linearized kinematic boundary condition Eq.~\ref{eq:nondim BVP flow BCs}
\begin{equation}
    \delta \dot{l}_\pm(t) = \frac{\partial_u \delta \sigma(u_\pm,t)}{L(\mathcal{A}+a)},
\end{equation}
yields the time evolution of the length perturbation
\begin{equation}
    \delta L(t) = \delta L_0 \exp \left(-\alpha t \right)
\end{equation}
with $\delta L_0 =\delta L(t=0)$ and $\alpha = \frac{2}{\mathcal{L} \sqrt{\mathcal{A}+a}} \tanh \left( \gamma/2 \right)$.
This describes an exponential decay toward zero and therefore also the stress perturbation in Eq.~\ref{eq:solution LSA stress} decays.
Thus, the evolution of $\delta \sigma$ is completely unrelated to $\delta a$, at least up to first order around homogeneous base states. In this sense, the adhesion density is a second order effect. 

Even when considering a base state with a non-uniform adhesion profile, like the ones found in figure Fig.~\ref{fig:Polarization mechanism}(g), a perturbation $\delta a$ would vanish in the equations Eq.~\ref{eq:nondim BVP stress} and Eq.~\ref{eq:nondim BVP flow BCs} in first order as long as the base state is uniform in the stress.

\clearpage

\section*{References}

\begin{thebibliography}{68}%
\makeatletter
\providecommand \@ifxundefined [1]{%
 \@ifx{#1\undefined}
}%
\providecommand \@ifnum [1]{%
 \ifnum #1\expandafter \@firstoftwo
 \else \expandafter \@secondoftwo
 \fi
}%
\providecommand \@ifx [1]{%
 \ifx #1\expandafter \@firstoftwo
 \else \expandafter \@secondoftwo
 \fi
}%
\providecommand \natexlab [1]{#1}%
\providecommand \enquote  [1]{``#1''}%
\providecommand \bibnamefont  [1]{#1}%
\providecommand \bibfnamefont [1]{#1}%
\providecommand \citenamefont [1]{#1}%
\providecommand \href@noop [0]{\@secondoftwo}%
\providecommand \href [0]{\begingroup \@sanitize@url \@href}%
\providecommand \@href[1]{\@@startlink{#1}\@@href}%
\providecommand \@@href[1]{\endgroup#1\@@endlink}%
\providecommand \@sanitize@url [0]{\catcode `\\12\catcode `\$12\catcode
  `\&12\catcode `\#12\catcode `\^12\catcode `\_12\catcode `\%12\relax}%
\providecommand \@@startlink[1]{}%
\providecommand \@@endlink[0]{}%
\providecommand \url  [0]{\begingroup\@sanitize@url \@url }%
\providecommand \@url [1]{\endgroup\@href {#1}{\urlprefix }}%
\providecommand \urlprefix  [0]{URL }%
\providecommand \Eprint [0]{\href }%
\providecommand \doibase [0]{http://dx.doi.org/}%
\providecommand \selectlanguage [0]{\@gobble}%
\providecommand \bibinfo  [0]{\@secondoftwo}%
\providecommand \bibfield  [0]{\@secondoftwo}%
\providecommand \translation [1]{[#1]}%
\providecommand \BibitemOpen [0]{}%
\providecommand \bibitemStop [0]{}%
\providecommand \bibitemNoStop [0]{.\EOS\space}%
\providecommand \EOS [0]{\spacefactor3000\relax}%
\providecommand \BibitemShut  [1]{\csname bibitem#1\endcsname}%
\let\auto@bib@innerbib\@empty
\bibitem [{\citenamefont {Bodor}\ \emph {et~al.}(2020)\citenamefont {Bodor},
  \citenamefont {Pönisch}, \citenamefont {Endres},\ and\ \citenamefont
  {Paluch}}]{bodor_cell_2020}%
  \BibitemOpen
  \bibfield  {author} {\bibinfo {author} {\bibfnamefont {D.~L.}\ \bibnamefont
  {Bodor}}, \bibinfo {author} {\bibfnamefont {W.}~\bibnamefont {Pönisch}},
  \bibinfo {author} {\bibfnamefont {R.~G.}\ \bibnamefont {Endres}}, \ and\
  \bibinfo {author} {\bibfnamefont {E.~K.}\ \bibnamefont {Paluch}},\ }\href
  {\doibase 10.1016/j.devcel.2020.02.013} {\bibfield  {journal} {\bibinfo
  {journal} {Developmental Cell}\ }\textbf {\bibinfo {volume} {52}},\ \bibinfo
  {pages} {550} (\bibinfo {year} {2020})}\BibitemShut {NoStop}%
\bibitem [{\citenamefont {Stuelten}\ \emph {et~al.}(2018)\citenamefont
  {Stuelten}, \citenamefont {Parent},\ and\ \citenamefont
  {Montell}}]{stuelten_cell_2018}%
  \BibitemOpen
  \bibfield  {author} {\bibinfo {author} {\bibfnamefont {C.~H.}\ \bibnamefont
  {Stuelten}}, \bibinfo {author} {\bibfnamefont {C.~A.}\ \bibnamefont
  {Parent}}, \ and\ \bibinfo {author} {\bibfnamefont {D.~J.}\ \bibnamefont
  {Montell}},\ }\href {\doibase 10.1038/nrc.2018.15} {\bibfield  {journal}
  {\bibinfo  {journal} {Nature Reviews Cancer}\ }\textbf {\bibinfo {volume}
  {18}},\ \bibinfo {pages} {296} (\bibinfo {year} {2018})},\ \bibinfo {note}
  {number: 5 Publisher: Nature Publishing Group}\BibitemShut {NoStop}%
\bibitem [{\citenamefont {Abercrombie}(1980)}]{abercrombie_croonian_1980}%
  \BibitemOpen
  \bibfield  {author} {\bibinfo {author} {\bibfnamefont {M.}~\bibnamefont
  {Abercrombie}},\ }\href {\doibase 10.1098/rspb.1980.0017} {\bibfield
  {journal} {\bibinfo  {journal} {Proceedings of the Royal Society of London.
  Series B. Biological Sciences}\ }\textbf {\bibinfo {volume} {207}},\ \bibinfo
  {pages} {129} (\bibinfo {year} {1980})},\ \bibinfo {note} {publisher: Royal
  Society}\BibitemShut {NoStop}%
\bibitem [{\citenamefont {Verkhovsky}\ \emph {et~al.}(1999)\citenamefont
  {Verkhovsky}, \citenamefont {Svitkina},\ and\ \citenamefont
  {Borisy}}]{verkhovsky_self-polarization_1999}%
  \BibitemOpen
  \bibfield  {author} {\bibinfo {author} {\bibfnamefont {A.~B.}\ \bibnamefont
  {Verkhovsky}}, \bibinfo {author} {\bibfnamefont {T.~M.}\ \bibnamefont
  {Svitkina}}, \ and\ \bibinfo {author} {\bibfnamefont {G.~G.}\ \bibnamefont
  {Borisy}},\ }\href {\doibase 10.1016/S0960-9822(99)80042-6} {\bibfield
  {journal} {\bibinfo  {journal} {Current Biology}\ }\textbf {\bibinfo {volume}
  {9}},\ \bibinfo {pages} {11} (\bibinfo {year} {1999})}\BibitemShut {NoStop}%
\bibitem [{\citenamefont {Ziebert}\ \emph {et~al.}(2012)\citenamefont
  {Ziebert}, \citenamefont {Swaminathan},\ and\ \citenamefont
  {Aranson}}]{Ziebert2012}%
  \BibitemOpen
  \bibfield  {author} {\bibinfo {author} {\bibfnamefont {F.}~\bibnamefont
  {Ziebert}}, \bibinfo {author} {\bibfnamefont {S.}~\bibnamefont
  {Swaminathan}}, \ and\ \bibinfo {author} {\bibfnamefont {I.~S.}\ \bibnamefont
  {Aranson}},\ }\href
  {https://royalsocietypublishing.org/doi/abs/10.1098/rsif.2011.0433}
  {\bibfield  {journal} {\bibinfo  {journal} {J. R. Soc. Interface}\ }\textbf
  {\bibinfo {volume} {9}},\ \bibinfo {pages} {1084} (\bibinfo {year}
  {2012})}\BibitemShut {NoStop}%
\bibitem [{\citenamefont {Lavi}\ \emph {et~al.}(2016)\citenamefont {Lavi},
  \citenamefont {Piel}, \citenamefont {Lennon-Duménil}, \citenamefont
  {Voituriez},\ and\ \citenamefont {Gov}}]{lavi_deterministic_2016}%
  \BibitemOpen
  \bibfield  {author} {\bibinfo {author} {\bibfnamefont {I.}~\bibnamefont
  {Lavi}}, \bibinfo {author} {\bibfnamefont {M.}~\bibnamefont {Piel}}, \bibinfo
  {author} {\bibfnamefont {A.-M.}\ \bibnamefont {Lennon-Duménil}}, \bibinfo
  {author} {\bibfnamefont {R.}~\bibnamefont {Voituriez}}, \ and\ \bibinfo
  {author} {\bibfnamefont {N.~S.}\ \bibnamefont {Gov}},\ }\href {\doibase
  10.1038/nphys3836} {\bibfield  {journal} {\bibinfo  {journal} {Nature
  Physics}\ }\textbf {\bibinfo {volume} {12}},\ \bibinfo {pages} {1146}
  (\bibinfo {year} {2016})},\ \bibinfo {note} {publisher: Nature Publishing
  Group}\BibitemShut {NoStop}%
\bibitem [{\citenamefont {Ron}\ \emph {et~al.}(2020)\citenamefont {Ron},
  \citenamefont {Monzo}, \citenamefont {Gauthier}, \citenamefont {Voituriez},\
  and\ \citenamefont {Gov}}]{ron_one-dimensional_2020}%
  \BibitemOpen
  \bibfield  {author} {\bibinfo {author} {\bibfnamefont {J.~E.}\ \bibnamefont
  {Ron}}, \bibinfo {author} {\bibfnamefont {P.}~\bibnamefont {Monzo}}, \bibinfo
  {author} {\bibfnamefont {N.~C.}\ \bibnamefont {Gauthier}}, \bibinfo {author}
  {\bibfnamefont {R.}~\bibnamefont {Voituriez}}, \ and\ \bibinfo {author}
  {\bibfnamefont {N.~S.}\ \bibnamefont {Gov}},\ }\href {\doibase
  10.1103/PhysRevResearch.2.033237} {\bibfield  {journal} {\bibinfo  {journal}
  {Physical Review Research}\ }\textbf {\bibinfo {volume} {2}},\ \bibinfo
  {pages} {033237} (\bibinfo {year} {2020})},\ \bibinfo {note} {publisher:
  American Physical Society}\BibitemShut {NoStop}%
\bibitem [{\citenamefont {Amiri}\ \emph {et~al.}(2023)\citenamefont {Amiri},
  \citenamefont {Heyn}, \citenamefont {Schreiber}, \citenamefont {Rädler},\
  and\ \citenamefont {Falcke}}]{amiri_multistability_2023}%
  \BibitemOpen
  \bibfield  {author} {\bibinfo {author} {\bibfnamefont {B.}~\bibnamefont
  {Amiri}}, \bibinfo {author} {\bibfnamefont {J.~C.~J.}\ \bibnamefont {Heyn}},
  \bibinfo {author} {\bibfnamefont {C.}~\bibnamefont {Schreiber}}, \bibinfo
  {author} {\bibfnamefont {J.~O.}\ \bibnamefont {Rädler}}, \ and\ \bibinfo
  {author} {\bibfnamefont {M.}~\bibnamefont {Falcke}},\ }\href {\doibase
  10.1016/j.bpj.2023.02.001} {\bibfield  {journal} {\bibinfo  {journal}
  {Biophysical Journal}\ }\textbf {\bibinfo {volume} {122}},\ \bibinfo {pages}
  {753} (\bibinfo {year} {2023})}\BibitemShut {NoStop}%
\bibitem [{\citenamefont {Maiuri}\ \emph {et~al.}(2012)\citenamefont {Maiuri},
  \citenamefont {Terriac}, \citenamefont {Paul-Gilloteaux}, \citenamefont
  {Vignaud}, \citenamefont {McNally}, \citenamefont {Onuffer}, \citenamefont
  {Thorn}, \citenamefont {Nguyen}, \citenamefont {Georgoulia}, \citenamefont
  {Soong}, \citenamefont {Jayo}, \citenamefont {Beil}, \citenamefont {Beneke},
  \citenamefont {Hong~Lim}, \citenamefont {Pei-Ying~Sim}, \citenamefont {Chu},
  \citenamefont {Jiménez-Dalmaroni}, \citenamefont {Joanny}, \citenamefont
  {Thiery}, \citenamefont {Erfle}, \citenamefont {Parsons}, \citenamefont
  {Mitchison}, \citenamefont {Lim}, \citenamefont {Lennon-Duménil},
  \citenamefont {Piel},\ and\ \citenamefont {Théry}}]{maiuri_first_2012}%
  \BibitemOpen
  \bibfield  {author} {\bibinfo {author} {\bibfnamefont {P.}~\bibnamefont
  {Maiuri}}, \bibinfo {author} {\bibfnamefont {E.}~\bibnamefont {Terriac}},
  \bibinfo {author} {\bibfnamefont {P.}~\bibnamefont {Paul-Gilloteaux}},
  \bibinfo {author} {\bibfnamefont {T.}~\bibnamefont {Vignaud}}, \bibinfo
  {author} {\bibfnamefont {K.}~\bibnamefont {McNally}}, \bibinfo {author}
  {\bibfnamefont {J.}~\bibnamefont {Onuffer}}, \bibinfo {author} {\bibfnamefont
  {K.}~\bibnamefont {Thorn}}, \bibinfo {author} {\bibfnamefont {P.~A.}\
  \bibnamefont {Nguyen}}, \bibinfo {author} {\bibfnamefont {N.}~\bibnamefont
  {Georgoulia}}, \bibinfo {author} {\bibfnamefont {D.}~\bibnamefont {Soong}},
  \bibinfo {author} {\bibfnamefont {A.}~\bibnamefont {Jayo}}, \bibinfo {author}
  {\bibfnamefont {N.}~\bibnamefont {Beil}}, \bibinfo {author} {\bibfnamefont
  {J.}~\bibnamefont {Beneke}}, \bibinfo {author} {\bibfnamefont {J.~C.}\
  \bibnamefont {Hong~Lim}}, \bibinfo {author} {\bibfnamefont {C.}~\bibnamefont
  {Pei-Ying~Sim}}, \bibinfo {author} {\bibfnamefont {Y.-S.}\ \bibnamefont
  {Chu}}, \bibinfo {author} {\bibfnamefont {A.}~\bibnamefont
  {Jiménez-Dalmaroni}}, \bibinfo {author} {\bibfnamefont {J.-F.}\ \bibnamefont
  {Joanny}}, \bibinfo {author} {\bibfnamefont {J.-P.}\ \bibnamefont {Thiery}},
  \bibinfo {author} {\bibfnamefont {H.}~\bibnamefont {Erfle}}, \bibinfo
  {author} {\bibfnamefont {M.}~\bibnamefont {Parsons}}, \bibinfo {author}
  {\bibfnamefont {T.~J.}\ \bibnamefont {Mitchison}}, \bibinfo {author}
  {\bibfnamefont {W.~A.}\ \bibnamefont {Lim}}, \bibinfo {author} {\bibfnamefont
  {A.-M.}\ \bibnamefont {Lennon-Duménil}}, \bibinfo {author} {\bibfnamefont
  {M.}~\bibnamefont {Piel}}, \ and\ \bibinfo {author} {\bibfnamefont
  {M.}~\bibnamefont {Théry}},\ }\href {\doibase 10.1016/j.cub.2012.07.052}
  {\bibfield  {journal} {\bibinfo  {journal} {Current Biology}\ }\textbf
  {\bibinfo {volume} {22}},\ \bibinfo {pages} {R673} (\bibinfo {year}
  {2012})}\BibitemShut {NoStop}%
\bibitem [{\citenamefont {Maiuri}\ \emph {et~al.}(2015)\citenamefont {Maiuri},
  \citenamefont {Rupprecht}, \citenamefont {Wieser}, \citenamefont {Ruprecht},
  \citenamefont {Bénichou}, \citenamefont {Carpi}, \citenamefont {Coppey},
  \citenamefont {De~Beco}, \citenamefont {Gov}, \citenamefont {Heisenberg},
  \citenamefont {Lage~Crespo}, \citenamefont {Lautenschlaeger}, \citenamefont
  {Le~Berre}, \citenamefont {Lennon-Dumenil}, \citenamefont {Raab},
  \citenamefont {Thiam}, \citenamefont {Piel}, \citenamefont {Sixt},\ and\
  \citenamefont {Voituriez}}]{maiuri_actin_2015}%
  \BibitemOpen
  \bibfield  {author} {\bibinfo {author} {\bibfnamefont {P.}~\bibnamefont
  {Maiuri}}, \bibinfo {author} {\bibfnamefont {J.-F.}\ \bibnamefont
  {Rupprecht}}, \bibinfo {author} {\bibfnamefont {S.}~\bibnamefont {Wieser}},
  \bibinfo {author} {\bibfnamefont {V.}~\bibnamefont {Ruprecht}}, \bibinfo
  {author} {\bibfnamefont {O.}~\bibnamefont {Bénichou}}, \bibinfo {author}
  {\bibfnamefont {N.}~\bibnamefont {Carpi}}, \bibinfo {author} {\bibfnamefont
  {M.}~\bibnamefont {Coppey}}, \bibinfo {author} {\bibfnamefont
  {S.}~\bibnamefont {De~Beco}}, \bibinfo {author} {\bibfnamefont
  {N.}~\bibnamefont {Gov}}, \bibinfo {author} {\bibfnamefont {C.-P.}\
  \bibnamefont {Heisenberg}}, \bibinfo {author} {\bibfnamefont
  {C.}~\bibnamefont {Lage~Crespo}}, \bibinfo {author} {\bibfnamefont
  {F.}~\bibnamefont {Lautenschlaeger}}, \bibinfo {author} {\bibfnamefont
  {M.}~\bibnamefont {Le~Berre}}, \bibinfo {author} {\bibfnamefont {A.-M.}\
  \bibnamefont {Lennon-Dumenil}}, \bibinfo {author} {\bibfnamefont
  {M.}~\bibnamefont {Raab}}, \bibinfo {author} {\bibfnamefont {H.-R.}\
  \bibnamefont {Thiam}}, \bibinfo {author} {\bibfnamefont {M.}~\bibnamefont
  {Piel}}, \bibinfo {author} {\bibfnamefont {M.}~\bibnamefont {Sixt}}, \ and\
  \bibinfo {author} {\bibfnamefont {R.}~\bibnamefont {Voituriez}},\ }\href
  {\doibase 10.1016/j.cell.2015.01.056} {\bibfield  {journal} {\bibinfo
  {journal} {Cell}\ }\textbf {\bibinfo {volume} {161}},\ \bibinfo {pages} {374}
  (\bibinfo {year} {2015})}\BibitemShut {NoStop}%
\bibitem [{\citenamefont {Hennig}\ \emph {et~al.}(2020)\citenamefont {Hennig},
  \citenamefont {Wang}, \citenamefont {Moreau}, \citenamefont {Valon},
  \citenamefont {DeBeco}, \citenamefont {Coppey}, \citenamefont {Miroshnikova},
  \citenamefont {Albiges-Rizo}, \citenamefont {Favard}, \citenamefont
  {Voituriez},\ and\ \citenamefont {Balland}}]{hennig_stick-slip_2020}%
  \BibitemOpen
  \bibfield  {author} {\bibinfo {author} {\bibfnamefont {K.}~\bibnamefont
  {Hennig}}, \bibinfo {author} {\bibfnamefont {I.}~\bibnamefont {Wang}},
  \bibinfo {author} {\bibfnamefont {P.}~\bibnamefont {Moreau}}, \bibinfo
  {author} {\bibfnamefont {L.}~\bibnamefont {Valon}}, \bibinfo {author}
  {\bibfnamefont {S.}~\bibnamefont {DeBeco}}, \bibinfo {author} {\bibfnamefont
  {M.}~\bibnamefont {Coppey}}, \bibinfo {author} {\bibfnamefont {Y.~A.}\
  \bibnamefont {Miroshnikova}}, \bibinfo {author} {\bibfnamefont
  {C.}~\bibnamefont {Albiges-Rizo}}, \bibinfo {author} {\bibfnamefont
  {C.}~\bibnamefont {Favard}}, \bibinfo {author} {\bibfnamefont
  {R.}~\bibnamefont {Voituriez}}, \ and\ \bibinfo {author} {\bibfnamefont
  {M.}~\bibnamefont {Balland}},\ }\href {\doibase 10.1126/sciadv.aau5670}
  {\bibfield  {journal} {\bibinfo  {journal} {Science Advances}\ }\textbf
  {\bibinfo {volume} {6}},\ \bibinfo {pages} {eaau5670} (\bibinfo {year}
  {2020})}\BibitemShut {NoStop}%
\bibitem [{\citenamefont {Schreiber}\ \emph {et~al.}(2021)\citenamefont
  {Schreiber}, \citenamefont {Amiri}, \citenamefont {Heyn}, \citenamefont
  {Rädler},\ and\ \citenamefont {Falcke}}]{schreiber_adhesionvelocity_2021}%
  \BibitemOpen
  \bibfield  {author} {\bibinfo {author} {\bibfnamefont {C.}~\bibnamefont
  {Schreiber}}, \bibinfo {author} {\bibfnamefont {B.}~\bibnamefont {Amiri}},
  \bibinfo {author} {\bibfnamefont {J.~C.~J.}\ \bibnamefont {Heyn}}, \bibinfo
  {author} {\bibfnamefont {J.~O.}\ \bibnamefont {Rädler}}, \ and\ \bibinfo
  {author} {\bibfnamefont {M.}~\bibnamefont {Falcke}},\ }\href {\doibase
  10.1073/pnas.2009959118} {\bibfield  {journal} {\bibinfo  {journal}
  {Proceedings of the National Academy of Sciences}\ }\textbf {\bibinfo
  {volume} {118}},\ \bibinfo {pages} {e2009959118} (\bibinfo {year} {2021})},\
  \bibinfo {note} {publisher: Proceedings of the National Academy of
  Sciences}\BibitemShut {NoStop}%
\bibitem [{\citenamefont {Kruse}\ \emph {et~al.}(2005)\citenamefont {Kruse},
  \citenamefont {Joanny}, \citenamefont {Jülicher}, \citenamefont {Prost},\
  and\ \citenamefont {Sekimoto}}]{kruse_generic_2005}%
  \BibitemOpen
  \bibfield  {author} {\bibinfo {author} {\bibfnamefont {K.}~\bibnamefont
  {Kruse}}, \bibinfo {author} {\bibfnamefont {J.~F.}\ \bibnamefont {Joanny}},
  \bibinfo {author} {\bibfnamefont {F.}~\bibnamefont {Jülicher}}, \bibinfo
  {author} {\bibfnamefont {J.}~\bibnamefont {Prost}}, \ and\ \bibinfo {author}
  {\bibfnamefont {K.}~\bibnamefont {Sekimoto}},\ }\href {\doibase
  10.1140/epje/e2005-00002-5} {\bibfield  {journal} {\bibinfo  {journal} {The
  European Physical Journal E}\ }\textbf {\bibinfo {volume} {16}},\ \bibinfo
  {pages} {5} (\bibinfo {year} {2005})}\BibitemShut {NoStop}%
\bibitem [{\citenamefont {Jülicher}\ \emph {et~al.}(2007)\citenamefont
  {Jülicher}, \citenamefont {Kruse}, \citenamefont {Prost},\ and\
  \citenamefont {Joanny}}]{julicher_active_2007}%
  \BibitemOpen
  \bibfield  {author} {\bibinfo {author} {\bibfnamefont {F.}~\bibnamefont
  {Jülicher}}, \bibinfo {author} {\bibfnamefont {K.}~\bibnamefont {Kruse}},
  \bibinfo {author} {\bibfnamefont {J.}~\bibnamefont {Prost}}, \ and\ \bibinfo
  {author} {\bibfnamefont {J.~F.}\ \bibnamefont {Joanny}},\ }\href {\doibase
  10.1016/j.physrep.2007.02.018} {\bibfield  {journal} {\bibinfo  {journal}
  {Physics Reports}\ }\bibinfo {series} {Nonequilibrium physics: {From} complex
  fluids to biological systems {III}. {Living} systems},\ \textbf {\bibinfo
  {volume} {449}},\ \bibinfo {pages} {3} (\bibinfo {year} {2007})}\BibitemShut
  {NoStop}%
\bibitem [{\citenamefont {Prost}\ \emph {et~al.}(2015)\citenamefont {Prost},
  \citenamefont {Jülicher},\ and\ \citenamefont {Joanny}}]{prost_active_2015}%
  \BibitemOpen
  \bibfield  {author} {\bibinfo {author} {\bibfnamefont {J.}~\bibnamefont
  {Prost}}, \bibinfo {author} {\bibfnamefont {F.}~\bibnamefont {Jülicher}}, \
  and\ \bibinfo {author} {\bibfnamefont {J.-F.}\ \bibnamefont {Joanny}},\
  }\href {\doibase 10.1038/nphys3224} {\bibfield  {journal} {\bibinfo
  {journal} {Nature Physics}\ }\textbf {\bibinfo {volume} {11}},\ \bibinfo
  {pages} {111} (\bibinfo {year} {2015})},\ \bibinfo {note} {number: 2
  Publisher: Nature Publishing Group}\BibitemShut {NoStop}%
\bibitem [{\citenamefont {Recho}\ \emph {et~al.}(2013)\citenamefont {Recho},
  \citenamefont {Putelat},\ and\ \citenamefont
  {Truskinovsky}}]{recho_contraction-driven_2013}%
  \BibitemOpen
  \bibfield  {author} {\bibinfo {author} {\bibfnamefont {P.}~\bibnamefont
  {Recho}}, \bibinfo {author} {\bibfnamefont {T.}~\bibnamefont {Putelat}}, \
  and\ \bibinfo {author} {\bibfnamefont {L.}~\bibnamefont {Truskinovsky}},\
  }\href {\doibase 10.1103/PhysRevLett.111.108102} {\bibfield  {journal}
  {\bibinfo  {journal} {Physical Review Letters}\ }\textbf {\bibinfo {volume}
  {111}},\ \bibinfo {pages} {108102} (\bibinfo {year} {2013})},\ \bibinfo
  {note} {publisher: American Physical Society}\BibitemShut {NoStop}%
\bibitem [{\citenamefont {Recho}\ \emph {et~al.}(2015)\citenamefont {Recho},
  \citenamefont {Putelat},\ and\ \citenamefont
  {Truskinovsky}}]{recho_mechanics_2015}%
  \BibitemOpen
  \bibfield  {author} {\bibinfo {author} {\bibfnamefont {P.}~\bibnamefont
  {Recho}}, \bibinfo {author} {\bibfnamefont {T.}~\bibnamefont {Putelat}}, \
  and\ \bibinfo {author} {\bibfnamefont {L.}~\bibnamefont {Truskinovsky}},\
  }\href {\doibase 10.1016/j.jmps.2015.08.006} {\bibfield  {journal} {\bibinfo
  {journal} {Journal of the Mechanics and Physics of Solids}\ }\textbf
  {\bibinfo {volume} {84}},\ \bibinfo {pages} {469} (\bibinfo {year}
  {2015})}\BibitemShut {NoStop}%
\bibitem [{\citenamefont {Wittmann}\ \emph {et~al.}(2020)\citenamefont
  {Wittmann}, \citenamefont {Dema},\ and\ \citenamefont {van
  Haren}}]{wittmann_lights_2020}%
  \BibitemOpen
  \bibfield  {author} {\bibinfo {author} {\bibfnamefont {T.}~\bibnamefont
  {Wittmann}}, \bibinfo {author} {\bibfnamefont {A.}~\bibnamefont {Dema}}, \
  and\ \bibinfo {author} {\bibfnamefont {J.}~\bibnamefont {van Haren}},\ }\href
  {\doibase 10.1016/j.ceb.2020.03.003} {\bibfield  {journal} {\bibinfo
  {journal} {Current Opinion in Cell Biology}\ }\bibinfo {series} {Cell
  {Dynamics}},\ \textbf {\bibinfo {volume} {66}},\ \bibinfo {pages} {1}
  (\bibinfo {year} {2020})}\BibitemShut {NoStop}%
\bibitem [{\citenamefont {Drozdowski}\ \emph {et~al.}(2021)\citenamefont
  {Drozdowski}, \citenamefont {Ziebert},\ and\ \citenamefont
  {Schwarz}}]{drozdowski_optogenetic_2021}%
  \BibitemOpen
  \bibfield  {author} {\bibinfo {author} {\bibfnamefont {O.~M.}\ \bibnamefont
  {Drozdowski}}, \bibinfo {author} {\bibfnamefont {F.}~\bibnamefont {Ziebert}},
  \ and\ \bibinfo {author} {\bibfnamefont {U.~S.}\ \bibnamefont {Schwarz}},\
  }\href {\doibase 10.1103/PhysRevE.104.024406} {\bibfield  {journal} {\bibinfo
   {journal} {Physical Review E}\ }\textbf {\bibinfo {volume} {104}},\ \bibinfo
  {pages} {024406} (\bibinfo {year} {2021})},\ \bibinfo {note} {publisher:
  American Physical Society}\BibitemShut {NoStop}%
\bibitem [{\citenamefont {Drozdowski}\ \emph {et~al.}(2023)\citenamefont
  {Drozdowski}, \citenamefont {Ziebert},\ and\ \citenamefont
  {Schwarz}}]{drozdowski_optogenetic_2023}%
  \BibitemOpen
  \bibfield  {author} {\bibinfo {author} {\bibfnamefont {O.~M.}\ \bibnamefont
  {Drozdowski}}, \bibinfo {author} {\bibfnamefont {F.}~\bibnamefont {Ziebert}},
  \ and\ \bibinfo {author} {\bibfnamefont {U.~S.}\ \bibnamefont {Schwarz}},\
  }\href {\doibase 10.1038/s42005-023-01275-0} {\bibfield  {journal} {\bibinfo
  {journal} {Communications Physics}\ }\textbf {\bibinfo {volume} {6}},\
  \bibinfo {pages} {1} (\bibinfo {year} {2023})},\ \bibinfo {note} {number: 1
  Publisher: Nature Publishing Group}\BibitemShut {NoStop}%
\bibitem [{\citenamefont {Humphrey}\ \emph {et~al.}(2014)\citenamefont
  {Humphrey}, \citenamefont {Dufresne},\ and\ \citenamefont
  {Schwartz}}]{humphrey_mechanotransduction_2014}%
  \BibitemOpen
  \bibfield  {author} {\bibinfo {author} {\bibfnamefont {J.~D.}\ \bibnamefont
  {Humphrey}}, \bibinfo {author} {\bibfnamefont {E.~R.}\ \bibnamefont
  {Dufresne}}, \ and\ \bibinfo {author} {\bibfnamefont {M.~A.}\ \bibnamefont
  {Schwartz}},\ }\href {\doibase 10.1038/nrm3896} {\bibfield  {journal}
  {\bibinfo  {journal} {Nature Reviews Molecular Cell Biology}\ }\textbf
  {\bibinfo {volume} {15}},\ \bibinfo {pages} {802} (\bibinfo {year} {2014})},\
  \bibinfo {note} {publisher: Nature Publishing Group}\BibitemShut {NoStop}%
\bibitem [{\citenamefont {Bell}(1978)}]{bell_models_1978}%
  \BibitemOpen
  \bibfield  {author} {\bibinfo {author} {\bibfnamefont {G.~I.}\ \bibnamefont
  {Bell}},\ }\href {\doibase 10.1126/science.347575} {\bibfield  {journal}
  {\bibinfo  {journal} {Science}\ }\textbf {\bibinfo {volume} {200}},\ \bibinfo
  {pages} {618} (\bibinfo {year} {1978})},\ \bibinfo {note} {publisher:
  American Association for the Advancement of Science}\BibitemShut {NoStop}%
\bibitem [{\citenamefont {Erdmann}\ and\ \citenamefont
  {Schwarz}(2004{\natexlab{a}})}]{erdmann_stability_2004}%
  \BibitemOpen
  \bibfield  {author} {\bibinfo {author} {\bibfnamefont {T.}~\bibnamefont
  {Erdmann}}\ and\ \bibinfo {author} {\bibfnamefont {U.~S.}\ \bibnamefont
  {Schwarz}},\ }\href {\doibase 10.1103/PhysRevLett.92.108102} {\bibfield
  {journal} {\bibinfo  {journal} {Physical Review Letters}\ }\textbf {\bibinfo
  {volume} {92}},\ \bibinfo {pages} {108102} (\bibinfo {year}
  {2004}{\natexlab{a}})},\ \bibinfo {note} {publisher: American Physical
  Society}\BibitemShut {NoStop}%
\bibitem [{\citenamefont {Sabass}\ and\ \citenamefont
  {Schwarz}(2010)}]{sabass_modeling_2010}%
  \BibitemOpen
  \bibfield  {author} {\bibinfo {author} {\bibfnamefont {B.}~\bibnamefont
  {Sabass}}\ and\ \bibinfo {author} {\bibfnamefont {U.~S.}\ \bibnamefont
  {Schwarz}},\ }\href {\doibase 10.1088/0953-8984/22/19/194112} {\bibfield
  {journal} {\bibinfo  {journal} {Journal of Physics: Condensed Matter}\
  }\textbf {\bibinfo {volume} {22}},\ \bibinfo {pages} {194112} (\bibinfo
  {year} {2010})}\BibitemShut {NoStop}%
\bibitem [{\citenamefont {Schwarz}\ and\ \citenamefont
  {Safran}(2013)}]{schwarz_physics_2013}%
  \BibitemOpen
  \bibfield  {author} {\bibinfo {author} {\bibfnamefont {U.~S.}\ \bibnamefont
  {Schwarz}}\ and\ \bibinfo {author} {\bibfnamefont {S.~A.}\ \bibnamefont
  {Safran}},\ }\href {\doibase 10.1103/RevModPhys.85.1327} {\bibfield
  {journal} {\bibinfo  {journal} {Reviews of Modern Physics}\ }\textbf
  {\bibinfo {volume} {85}},\ \bibinfo {pages} {1327} (\bibinfo {year}
  {2013})},\ \bibinfo {note} {publisher: American Physical Society}\BibitemShut
  {NoStop}%
\bibitem [{\citenamefont {Schallamach}(1963)}]{schallamach_theory_1963}%
  \BibitemOpen
  \bibfield  {author} {\bibinfo {author} {\bibfnamefont {A.}~\bibnamefont
  {Schallamach}},\ }\href {\doibase 10.1016/0043-1648(63)90206-0} {\bibfield
  {journal} {\bibinfo  {journal} {Wear}\ }\textbf {\bibinfo {volume} {6}},\
  \bibinfo {pages} {375} (\bibinfo {year} {1963})}\BibitemShut {NoStop}%
\bibitem [{\citenamefont {Filippov}\ \emph {et~al.}(2004)\citenamefont
  {Filippov}, \citenamefont {Klafter},\ and\ \citenamefont
  {Urbakh}}]{filippov_friction_2004}%
  \BibitemOpen
  \bibfield  {author} {\bibinfo {author} {\bibfnamefont {A.~E.}\ \bibnamefont
  {Filippov}}, \bibinfo {author} {\bibfnamefont {J.}~\bibnamefont {Klafter}}, \
  and\ \bibinfo {author} {\bibfnamefont {M.}~\bibnamefont {Urbakh}},\ }\href
  {\doibase 10.1103/PhysRevLett.92.135503} {\bibfield  {journal} {\bibinfo
  {journal} {Physical Review Letters}\ }\textbf {\bibinfo {volume} {92}},\
  \bibinfo {pages} {135503} (\bibinfo {year} {2004})},\ \bibinfo {note}
  {publisher: American Physical Society}\BibitemShut {NoStop}%
\bibitem [{\citenamefont {Sens}(2020)}]{sens_stickslip_2020}%
  \BibitemOpen
  \bibfield  {author} {\bibinfo {author} {\bibfnamefont {P.}~\bibnamefont
  {Sens}},\ }\href {\doibase 10.1073/pnas.2011785117} {\bibfield  {journal}
  {\bibinfo  {journal} {Proceedings of the National Academy of Sciences}\
  }\textbf {\bibinfo {volume} {117}},\ \bibinfo {pages} {24670} (\bibinfo
  {year} {2020})},\ \bibinfo {note} {publisher: Proceedings of the National
  Academy of Sciences}\BibitemShut {NoStop}%
\bibitem [{\citenamefont {Chan}\ and\ \citenamefont
  {Odde}(2008)}]{chan_traction_2008}%
  \BibitemOpen
  \bibfield  {author} {\bibinfo {author} {\bibfnamefont {C.~E.}\ \bibnamefont
  {Chan}}\ and\ \bibinfo {author} {\bibfnamefont {D.~J.}\ \bibnamefont
  {Odde}},\ }\href {\doibase 10.1126/science.1163595} {\bibfield  {journal}
  {\bibinfo  {journal} {Science}\ }\textbf {\bibinfo {volume} {322}},\ \bibinfo
  {pages} {1687} (\bibinfo {year} {2008})},\ \bibinfo {note} {publisher:
  American Association for the Advancement of Science}\BibitemShut {NoStop}%
\bibitem [{\citenamefont {Bangasser}\ \emph {et~al.}(2013)\citenamefont
  {Bangasser}, \citenamefont {Rosenfeld},\ and\ \citenamefont
  {Odde}}]{bangasser_determinants_2013}%
  \BibitemOpen
  \bibfield  {author} {\bibinfo {author} {\bibfnamefont {B.}~\bibnamefont
  {Bangasser}}, \bibinfo {author} {\bibfnamefont {S.}~\bibnamefont
  {Rosenfeld}}, \ and\ \bibinfo {author} {\bibfnamefont {D.}~\bibnamefont
  {Odde}},\ }\href {\doibase 10.1016/j.bpj.2013.06.027} {\bibfield  {journal}
  {\bibinfo  {journal} {Biophysical Journal}\ }\textbf {\bibinfo {volume}
  {105}},\ \bibinfo {pages} {581} (\bibinfo {year} {2013})}\BibitemShut
  {NoStop}%
\bibitem [{\citenamefont {Oria}\ \emph {et~al.}(2017)\citenamefont {Oria},
  \citenamefont {Wiegand}, \citenamefont {Escribano}, \citenamefont
  {Elosegui-Artola}, \citenamefont {Uriarte}, \citenamefont {Moreno-Pulido},
  \citenamefont {Platzman}, \citenamefont {Delcanale}, \citenamefont
  {Albertazzi}, \citenamefont {Navajas}, \citenamefont {Trepat}, \citenamefont
  {García-Aznar}, \citenamefont {Cavalcanti-Adam},\ and\ \citenamefont
  {Roca-Cusachs}}]{oria_force_2017}%
  \BibitemOpen
  \bibfield  {author} {\bibinfo {author} {\bibfnamefont {R.}~\bibnamefont
  {Oria}}, \bibinfo {author} {\bibfnamefont {T.}~\bibnamefont {Wiegand}},
  \bibinfo {author} {\bibfnamefont {J.}~\bibnamefont {Escribano}}, \bibinfo
  {author} {\bibfnamefont {A.}~\bibnamefont {Elosegui-Artola}}, \bibinfo
  {author} {\bibfnamefont {J.~J.}\ \bibnamefont {Uriarte}}, \bibinfo {author}
  {\bibfnamefont {C.}~\bibnamefont {Moreno-Pulido}}, \bibinfo {author}
  {\bibfnamefont {I.}~\bibnamefont {Platzman}}, \bibinfo {author}
  {\bibfnamefont {P.}~\bibnamefont {Delcanale}}, \bibinfo {author}
  {\bibfnamefont {L.}~\bibnamefont {Albertazzi}}, \bibinfo {author}
  {\bibfnamefont {D.}~\bibnamefont {Navajas}}, \bibinfo {author} {\bibfnamefont
  {X.}~\bibnamefont {Trepat}}, \bibinfo {author} {\bibfnamefont {J.~M.}\
  \bibnamefont {García-Aznar}}, \bibinfo {author} {\bibfnamefont {E.~A.}\
  \bibnamefont {Cavalcanti-Adam}}, \ and\ \bibinfo {author} {\bibfnamefont
  {P.}~\bibnamefont {Roca-Cusachs}},\ }\href {\doibase 10.1038/nature24662}
  {\bibfield  {journal} {\bibinfo  {journal} {Nature}\ }\textbf {\bibinfo
  {volume} {552}},\ \bibinfo {pages} {219} (\bibinfo {year} {2017})},\ \bibinfo
  {note} {publisher: Nature Publishing Group}\BibitemShut {NoStop}%
\bibitem [{\citenamefont {Elosegui-Artola}\ \emph {et~al.}(2018)\citenamefont
  {Elosegui-Artola}, \citenamefont {Trepat},\ and\ \citenamefont
  {Roca-Cusachs}}]{elosegui-artola_control_2018}%
  \BibitemOpen
  \bibfield  {author} {\bibinfo {author} {\bibfnamefont {A.}~\bibnamefont
  {Elosegui-Artola}}, \bibinfo {author} {\bibfnamefont {X.}~\bibnamefont
  {Trepat}}, \ and\ \bibinfo {author} {\bibfnamefont {P.}~\bibnamefont
  {Roca-Cusachs}},\ }\href {\doibase 10.1016/j.tcb.2018.01.008} {\bibfield
  {journal} {\bibinfo  {journal} {Trends in Cell Biology}\ }\textbf {\bibinfo
  {volume} {28}},\ \bibinfo {pages} {356} (\bibinfo {year} {2018})}\BibitemShut
  {NoStop}%
\bibitem [{\citenamefont {Alonso-Matilla}\ \emph {et~al.}(2023)\citenamefont
  {Alonso-Matilla}, \citenamefont {Provenzano},\ and\ \citenamefont
  {Odde}}]{alonso-matilla_optimal_2023}%
  \BibitemOpen
  \bibfield  {author} {\bibinfo {author} {\bibfnamefont {R.}~\bibnamefont
  {Alonso-Matilla}}, \bibinfo {author} {\bibfnamefont {P.~P.}\ \bibnamefont
  {Provenzano}}, \ and\ \bibinfo {author} {\bibfnamefont {D.~J.}\ \bibnamefont
  {Odde}},\ }\href {\doibase 10.1016/j.bpj.2023.07.012} {\bibfield  {journal}
  {\bibinfo  {journal} {Biophysical Journal}\ }\textbf {\bibinfo {volume}
  {122}},\ \bibinfo {pages} {3369} (\bibinfo {year} {2023})}\BibitemShut
  {NoStop}%
\bibitem [{\citenamefont {Lo}\ \emph {et~al.}(2024)\citenamefont {Lo},
  \citenamefont {Tseng},\ and\ \citenamefont
  {Chen}}]{lo_mechanosensitive_2024}%
  \BibitemOpen
  \bibfield  {author} {\bibinfo {author} {\bibfnamefont {J.-Y.}\ \bibnamefont
  {Lo}}, \bibinfo {author} {\bibfnamefont {Y.-H.}\ \bibnamefont {Tseng}}, \
  and\ \bibinfo {author} {\bibfnamefont {H.-Y.}\ \bibnamefont {Chen}},\ }\href
  {\doibase 10.1103/PhysRevResearch.6.013164} {\bibfield  {journal} {\bibinfo
  {journal} {Physical Review Research}\ }\textbf {\bibinfo {volume} {6}},\
  \bibinfo {pages} {013164} (\bibinfo {year} {2024})},\ \bibinfo {note}
  {publisher: American Physical Society}\BibitemShut {NoStop}%
\bibitem [{\citenamefont {Pollard}\ and\ \citenamefont
  {Borisy}(2003)}]{pollard_cellular_2003}%
  \BibitemOpen
  \bibfield  {author} {\bibinfo {author} {\bibfnamefont {T.~D.}\ \bibnamefont
  {Pollard}}\ and\ \bibinfo {author} {\bibfnamefont {G.~G.}\ \bibnamefont
  {Borisy}},\ }\href {\doibase 10.1016/s0092-8674(03)00120-x} {\bibfield
  {journal} {\bibinfo  {journal} {Cell}\ }\textbf {\bibinfo {volume} {112}},\
  \bibinfo {pages} {453} (\bibinfo {year} {2003})}\BibitemShut {NoStop}%
\bibitem [{\citenamefont {Ridley}\ \emph {et~al.}(2003)\citenamefont {Ridley},
  \citenamefont {Schwartz}, \citenamefont {Burridge}, \citenamefont {Firtel},
  \citenamefont {Ginsberg}, \citenamefont {Borisy}, \citenamefont {Parsons},\
  and\ \citenamefont {Horwitz}}]{ridley_cell_2003}%
  \BibitemOpen
  \bibfield  {author} {\bibinfo {author} {\bibfnamefont {A.~J.}\ \bibnamefont
  {Ridley}}, \bibinfo {author} {\bibfnamefont {M.~A.}\ \bibnamefont
  {Schwartz}}, \bibinfo {author} {\bibfnamefont {K.}~\bibnamefont {Burridge}},
  \bibinfo {author} {\bibfnamefont {R.~A.}\ \bibnamefont {Firtel}}, \bibinfo
  {author} {\bibfnamefont {M.~H.}\ \bibnamefont {Ginsberg}}, \bibinfo {author}
  {\bibfnamefont {G.}~\bibnamefont {Borisy}}, \bibinfo {author} {\bibfnamefont
  {J.~T.}\ \bibnamefont {Parsons}}, \ and\ \bibinfo {author} {\bibfnamefont
  {A.~R.}\ \bibnamefont {Horwitz}},\ }\href {\doibase 10.1126/science.1092053}
  {\bibfield  {journal} {\bibinfo  {journal} {Science (New York, N.Y.)}\
  }\textbf {\bibinfo {volume} {302}},\ \bibinfo {pages} {1704} (\bibinfo {year}
  {2003})}\BibitemShut {NoStop}%
\bibitem [{\citenamefont {Cramer}(2010)}]{cramer_forming_2010}%
  \BibitemOpen
  \bibfield  {author} {\bibinfo {author} {\bibfnamefont {L.~P.}\ \bibnamefont
  {Cramer}},\ }\href {\doibase 10.1038/ncb0710-628} {\bibfield  {journal}
  {\bibinfo  {journal} {Nature Cell Biology}\ }\textbf {\bibinfo {volume}
  {12}},\ \bibinfo {pages} {628} (\bibinfo {year} {2010})},\ \bibinfo {note}
  {number: 7 Publisher: Nature Publishing Group}\BibitemShut {NoStop}%
\bibitem [{\citenamefont {Yam}\ \emph {et~al.}(2007)\citenamefont {Yam},
  \citenamefont {Wilson}, \citenamefont {Ji}, \citenamefont {Hebert},
  \citenamefont {Barnhart}, \citenamefont {Dye}, \citenamefont {Wiseman},
  \citenamefont {Danuser},\ and\ \citenamefont
  {Theriot}}]{yam_actinmyosin_2007}%
  \BibitemOpen
  \bibfield  {author} {\bibinfo {author} {\bibfnamefont {P.~T.}\ \bibnamefont
  {Yam}}, \bibinfo {author} {\bibfnamefont {C.~A.}\ \bibnamefont {Wilson}},
  \bibinfo {author} {\bibfnamefont {L.}~\bibnamefont {Ji}}, \bibinfo {author}
  {\bibfnamefont {B.}~\bibnamefont {Hebert}}, \bibinfo {author} {\bibfnamefont
  {E.~L.}\ \bibnamefont {Barnhart}}, \bibinfo {author} {\bibfnamefont {N.~A.}\
  \bibnamefont {Dye}}, \bibinfo {author} {\bibfnamefont {P.~W.}\ \bibnamefont
  {Wiseman}}, \bibinfo {author} {\bibfnamefont {G.}~\bibnamefont {Danuser}}, \
  and\ \bibinfo {author} {\bibfnamefont {J.~A.}\ \bibnamefont {Theriot}},\
  }\href {\doibase 10.1083/jcb.200706012} {\bibfield  {journal} {\bibinfo
  {journal} {The Journal of Cell Biology}\ }\textbf {\bibinfo {volume} {178}},\
  \bibinfo {pages} {1207} (\bibinfo {year} {2007})}\BibitemShut {NoStop}%
\bibitem [{\citenamefont {Barnhart}\ \emph {et~al.}(2015)\citenamefont
  {Barnhart}, \citenamefont {Lee}, \citenamefont {Allen}, \citenamefont
  {Theriot},\ and\ \citenamefont {Mogilner}}]{barnhart_balance_2015}%
  \BibitemOpen
  \bibfield  {author} {\bibinfo {author} {\bibfnamefont {E.}~\bibnamefont
  {Barnhart}}, \bibinfo {author} {\bibfnamefont {K.-C.}\ \bibnamefont {Lee}},
  \bibinfo {author} {\bibfnamefont {G.~M.}\ \bibnamefont {Allen}}, \bibinfo
  {author} {\bibfnamefont {J.~A.}\ \bibnamefont {Theriot}}, \ and\ \bibinfo
  {author} {\bibfnamefont {A.}~\bibnamefont {Mogilner}},\ }\href {\doibase
  10.1073/pnas.1417257112} {\bibfield  {journal} {\bibinfo  {journal}
  {Proceedings of the National Academy of Sciences}\ }\textbf {\bibinfo
  {volume} {112}},\ \bibinfo {pages} {5045} (\bibinfo {year} {2015})},\
  \bibinfo {note} {publisher: Proceedings of the National Academy of
  Sciences}\BibitemShut {NoStop}%
\bibitem [{\citenamefont {Ricoult}\ \emph {et~al.}(2015)\citenamefont
  {Ricoult}, \citenamefont {Kennedy},\ and\ \citenamefont
  {Juncker}}]{ricoult_substrate-bound_2015}%
  \BibitemOpen
  \bibfield  {author} {\bibinfo {author} {\bibfnamefont {S.~G.}\ \bibnamefont
  {Ricoult}}, \bibinfo {author} {\bibfnamefont {T.~E.}\ \bibnamefont
  {Kennedy}}, \ and\ \bibinfo {author} {\bibfnamefont {D.}~\bibnamefont
  {Juncker}},\ }\href
  {https://www.frontiersin.org/articles/10.3389/fbioe.2015.00040} {\bibfield
  {journal} {\bibinfo  {journal} {Frontiers in Bioengineering and
  Biotechnology}\ }\textbf {\bibinfo {volume} {3}} (\bibinfo {year}
  {2015})}\BibitemShut {NoStop}%
\bibitem [{\citenamefont {Giannone}\ \emph {et~al.}(2009)\citenamefont
  {Giannone}, \citenamefont {Mège},\ and\ \citenamefont
  {Thoumine}}]{giannone_multi-level_2009}%
  \BibitemOpen
  \bibfield  {author} {\bibinfo {author} {\bibfnamefont {G.}~\bibnamefont
  {Giannone}}, \bibinfo {author} {\bibfnamefont {R.-M.}\ \bibnamefont {Mège}},
  \ and\ \bibinfo {author} {\bibfnamefont {O.}~\bibnamefont {Thoumine}},\
  }\href {\doibase 10.1016/j.tcb.2009.07.001} {\bibfield  {journal} {\bibinfo
  {journal} {Trends in Cell Biology}\ }\textbf {\bibinfo {volume} {19}},\
  \bibinfo {pages} {475} (\bibinfo {year} {2009})}\BibitemShut {NoStop}%
\bibitem [{\citenamefont {Tawada}\ and\ \citenamefont
  {Sekimoto}(1991)}]{tawada1991protein}%
  \BibitemOpen
  \bibfield  {author} {\bibinfo {author} {\bibfnamefont {K.}~\bibnamefont
  {Tawada}}\ and\ \bibinfo {author} {\bibfnamefont {K.}~\bibnamefont
  {Sekimoto}},\ }\href@noop {} {\bibfield  {journal} {\bibinfo  {journal}
  {Journal of theoretical biology}\ }\textbf {\bibinfo {volume} {150}},\
  \bibinfo {pages} {193} (\bibinfo {year} {1991})}\BibitemShut {NoStop}%
\bibitem [{\citenamefont {Barnhart}\ \emph {et~al.}(2011)\citenamefont
  {Barnhart}, \citenamefont {Lee}, \citenamefont {Keren}, \citenamefont
  {Mogilner},\ and\ \citenamefont
  {Theriot}}]{barnhart_adhesion-dependent_2011}%
  \BibitemOpen
  \bibfield  {author} {\bibinfo {author} {\bibfnamefont {E.~L.}\ \bibnamefont
  {Barnhart}}, \bibinfo {author} {\bibfnamefont {K.-C.}\ \bibnamefont {Lee}},
  \bibinfo {author} {\bibfnamefont {K.}~\bibnamefont {Keren}}, \bibinfo
  {author} {\bibfnamefont {A.}~\bibnamefont {Mogilner}}, \ and\ \bibinfo
  {author} {\bibfnamefont {J.~A.}\ \bibnamefont {Theriot}},\ }\href {\doibase
  10.1371/journal.pbio.1001059} {\bibfield  {journal} {\bibinfo  {journal}
  {PLOS Biology}\ }\textbf {\bibinfo {volume} {9}},\ \bibinfo {pages}
  {e1001059} (\bibinfo {year} {2011})},\ \bibinfo {note} {publisher: Public
  Library of Science}\BibitemShut {NoStop}%
\bibitem [{\citenamefont {Löber}\ \emph {et~al.}(2014)\citenamefont {Löber},
  \citenamefont {Ziebert},\ and\ \citenamefont {Aranson}}]{LoeberSM}%
  \BibitemOpen
  \bibfield  {author} {\bibinfo {author} {\bibfnamefont {J.}~\bibnamefont
  {Löber}}, \bibinfo {author} {\bibfnamefont {F.}~\bibnamefont {Ziebert}}, \
  and\ \bibinfo {author} {\bibfnamefont {I.~S.}\ \bibnamefont {Aranson}},\
  }\href {https://pubs.rsc.org/en/content/articlelanding/2014/sm/c3sm51597d}
  {\bibfield  {journal} {\bibinfo  {journal} {Soft Matter}\ }\textbf {\bibinfo
  {volume} {10}},\ \bibinfo {pages} {1365} (\bibinfo {year}
  {2014})}\BibitemShut {NoStop}%
\bibitem [{\citenamefont {Spiess}\ \emph {et~al.}(2018)\citenamefont {Spiess},
  \citenamefont {Hernandez-Varas}, \citenamefont {Oddone}, \citenamefont
  {Olofsson}, \citenamefont {Blom}, \citenamefont {Waithe}, \citenamefont
  {Lock}, \citenamefont {Lakadamyali},\ and\ \citenamefont
  {Strömblad}}]{spiess_active_2018}%
  \BibitemOpen
  \bibfield  {author} {\bibinfo {author} {\bibfnamefont {M.}~\bibnamefont
  {Spiess}}, \bibinfo {author} {\bibfnamefont {P.}~\bibnamefont
  {Hernandez-Varas}}, \bibinfo {author} {\bibfnamefont {A.}~\bibnamefont
  {Oddone}}, \bibinfo {author} {\bibfnamefont {H.}~\bibnamefont {Olofsson}},
  \bibinfo {author} {\bibfnamefont {H.}~\bibnamefont {Blom}}, \bibinfo {author}
  {\bibfnamefont {D.}~\bibnamefont {Waithe}}, \bibinfo {author} {\bibfnamefont
  {J.~G.}\ \bibnamefont {Lock}}, \bibinfo {author} {\bibfnamefont
  {M.}~\bibnamefont {Lakadamyali}}, \ and\ \bibinfo {author} {\bibfnamefont
  {S.}~\bibnamefont {Strömblad}},\ }\href {\doibase 10.1083/jcb.201707075}
  {\bibfield  {journal} {\bibinfo  {journal} {Journal of Cell Biology}\
  }\textbf {\bibinfo {volume} {217}},\ \bibinfo {pages} {1929} (\bibinfo {year}
  {2018})}\BibitemShut {NoStop}%
\bibitem [{\citenamefont {Janeš}\ \emph {et~al.}(2022)\citenamefont {Janeš},
  \citenamefont {Monzel}, \citenamefont {Schmidt}, \citenamefont {Merkel},
  \citenamefont {Seifert}, \citenamefont {Sengupta},\ and\ \citenamefont
  {Smith}}]{janes_first-principle_2022}%
  \BibitemOpen
  \bibfield  {author} {\bibinfo {author} {\bibfnamefont {J.~A.}\ \bibnamefont
  {Janeš}}, \bibinfo {author} {\bibfnamefont {C.}~\bibnamefont {Monzel}},
  \bibinfo {author} {\bibfnamefont {D.}~\bibnamefont {Schmidt}}, \bibinfo
  {author} {\bibfnamefont {R.}~\bibnamefont {Merkel}}, \bibinfo {author}
  {\bibfnamefont {U.}~\bibnamefont {Seifert}}, \bibinfo {author} {\bibfnamefont
  {K.}~\bibnamefont {Sengupta}}, \ and\ \bibinfo {author} {\bibfnamefont
  {A.-S.}\ \bibnamefont {Smith}},\ }\href {\doibase 10.1103/PhysRevX.12.031030}
  {\bibfield  {journal} {\bibinfo  {journal} {Physical Review X}\ }\textbf
  {\bibinfo {volume} {12}},\ \bibinfo {pages} {031030} (\bibinfo {year}
  {2022})},\ \bibinfo {note} {publisher: American Physical Society}\BibitemShut
  {NoStop}%
\bibitem [{\citenamefont {Monzel}\ \emph {et~al.}(2015)\citenamefont {Monzel},
  \citenamefont {Schmidt}, \citenamefont {Kleusch}, \citenamefont
  {Kirchenb{\"u}chler}, \citenamefont {Seifert}, \citenamefont {Smith},
  \citenamefont {Sengupta},\ and\ \citenamefont
  {Merkel}}]{monzel2015measuring}%
  \BibitemOpen
  \bibfield  {author} {\bibinfo {author} {\bibfnamefont {C.}~\bibnamefont
  {Monzel}}, \bibinfo {author} {\bibfnamefont {D.}~\bibnamefont {Schmidt}},
  \bibinfo {author} {\bibfnamefont {C.}~\bibnamefont {Kleusch}}, \bibinfo
  {author} {\bibfnamefont {D.}~\bibnamefont {Kirchenb{\"u}chler}}, \bibinfo
  {author} {\bibfnamefont {U.}~\bibnamefont {Seifert}}, \bibinfo {author}
  {\bibfnamefont {A.-S.}\ \bibnamefont {Smith}}, \bibinfo {author}
  {\bibfnamefont {K.}~\bibnamefont {Sengupta}}, \ and\ \bibinfo {author}
  {\bibfnamefont {R.}~\bibnamefont {Merkel}},\ }\href@noop {} {\bibfield
  {journal} {\bibinfo  {journal} {Nature communications}\ }\textbf {\bibinfo
  {volume} {6}},\ \bibinfo {pages} {8162} (\bibinfo {year} {2015})}\BibitemShut
  {NoStop}%
\bibitem [{\citenamefont {Erdmann}\ and\ \citenamefont
  {Schwarz}(2004{\natexlab{b}})}]{erdmann_stochastic_2004}%
  \BibitemOpen
  \bibfield  {author} {\bibinfo {author} {\bibfnamefont {T.}~\bibnamefont
  {Erdmann}}\ and\ \bibinfo {author} {\bibfnamefont {U.~S.}\ \bibnamefont
  {Schwarz}},\ }\href {\doibase 10.1063/1.1805496} {\bibfield  {journal}
  {\bibinfo  {journal} {The Journal of Chemical Physics}\ }\textbf {\bibinfo
  {volume} {121}},\ \bibinfo {pages} {8997} (\bibinfo {year}
  {2004}{\natexlab{b}})}\BibitemShut {NoStop}%
\bibitem [{\citenamefont {Wiseman}\ \emph {et~al.}(2004)\citenamefont
  {Wiseman}, \citenamefont {Brown}, \citenamefont {Webb}, \citenamefont
  {Hebert}, \citenamefont {Johnson}, \citenamefont {Squier}, \citenamefont
  {Ellisman},\ and\ \citenamefont {Horwitz}}]{wiseman2004spatial}%
  \BibitemOpen
  \bibfield  {author} {\bibinfo {author} {\bibfnamefont {P.~W.}\ \bibnamefont
  {Wiseman}}, \bibinfo {author} {\bibfnamefont {C.~M.}\ \bibnamefont {Brown}},
  \bibinfo {author} {\bibfnamefont {D.~J.}\ \bibnamefont {Webb}}, \bibinfo
  {author} {\bibfnamefont {B.}~\bibnamefont {Hebert}}, \bibinfo {author}
  {\bibfnamefont {N.~L.}\ \bibnamefont {Johnson}}, \bibinfo {author}
  {\bibfnamefont {J.~A.}\ \bibnamefont {Squier}}, \bibinfo {author}
  {\bibfnamefont {M.~H.}\ \bibnamefont {Ellisman}}, \ and\ \bibinfo {author}
  {\bibfnamefont {A.}~\bibnamefont {Horwitz}},\ }\href@noop {} {\bibfield
  {journal} {\bibinfo  {journal} {Journal of cell science}\ }\textbf {\bibinfo
  {volume} {117}},\ \bibinfo {pages} {5521} (\bibinfo {year}
  {2004})}\BibitemShut {NoStop}%
\bibitem [{\citenamefont {Zaidel-Bar}\ \emph {et~al.}(2004)\citenamefont
  {Zaidel-Bar}, \citenamefont {Cohen}, \citenamefont {Addadi},\ and\
  \citenamefont {Geiger}}]{zaidel-bar_hierarchical_2004}%
  \BibitemOpen
  \bibfield  {author} {\bibinfo {author} {\bibfnamefont {R.}~\bibnamefont
  {Zaidel-Bar}}, \bibinfo {author} {\bibfnamefont {M.}~\bibnamefont {Cohen}},
  \bibinfo {author} {\bibfnamefont {L.}~\bibnamefont {Addadi}}, \ and\ \bibinfo
  {author} {\bibfnamefont {B.}~\bibnamefont {Geiger}},\ }\href {\doibase
  10.1042/bst0320416} {\bibfield  {journal} {\bibinfo  {journal} {Biochemical
  Society Transactions}\ }\textbf {\bibinfo {volume} {32}},\ \bibinfo {pages}
  {416} (\bibinfo {year} {2004})}\BibitemShut {NoStop}%
\bibitem [{\citenamefont {Doedel}\ \emph {et~al.}(2007)\citenamefont {Doedel},
  \citenamefont {Champneys}, \citenamefont {Dercole}, \citenamefont
  {Fairgrieve}, \citenamefont {Kuznetsov}, \citenamefont {Oldeman},
  \citenamefont {Paffenroth}, \citenamefont {Sandstede}, \citenamefont {Wang},\
  and\ \citenamefont {Zhang}}]{Doedel_2007_AUTO07p}%
  \BibitemOpen
  \bibfield  {author} {\bibinfo {author} {\bibfnamefont {E.~J.}\ \bibnamefont
  {Doedel}}, \bibinfo {author} {\bibfnamefont {A.~R.}\ \bibnamefont
  {Champneys}}, \bibinfo {author} {\bibfnamefont {F.}~\bibnamefont {Dercole}},
  \bibinfo {author} {\bibfnamefont {T.~F.}\ \bibnamefont {Fairgrieve}},
  \bibinfo {author} {\bibfnamefont {Y.~A.}\ \bibnamefont {Kuznetsov}}, \bibinfo
  {author} {\bibfnamefont {B.}~\bibnamefont {Oldeman}}, \bibinfo {author}
  {\bibfnamefont {R.}~\bibnamefont {Paffenroth}}, \bibinfo {author}
  {\bibfnamefont {B.}~\bibnamefont {Sandstede}}, \bibinfo {author}
  {\bibfnamefont {X.}~\bibnamefont {Wang}}, \ and\ \bibinfo {author}
  {\bibfnamefont {C.}~\bibnamefont {Zhang}},\ }\href@noop {} {\enquote
  {\bibinfo {title} {{AUTO-07P}: Continuation and bifurcation software for
  ordinary differential equations},}\ } (\bibinfo {year} {2007})\BibitemShut
  {NoStop}%
\bibitem [{\citenamefont {Guyer}\ \emph {et~al.}(2009)\citenamefont {Guyer},
  \citenamefont {Wheeler},\ and\ \citenamefont
  {Warren}}]{Guyer_2009_CompSciEng_FiPy}%
  \BibitemOpen
  \bibfield  {author} {\bibinfo {author} {\bibfnamefont {J.~E.}\ \bibnamefont
  {Guyer}}, \bibinfo {author} {\bibfnamefont {D.}~\bibnamefont {Wheeler}}, \
  and\ \bibinfo {author} {\bibfnamefont {J.~A.}\ \bibnamefont {Warren}},\
  }\href {\doibase 10.1109/MCSE.2009.52} {\bibfield  {journal} {\bibinfo
  {journal} {Computing in Science \& Engineering}\ }\textbf {\bibinfo {volume}
  {11}},\ \bibinfo {pages} {6} (\bibinfo {year} {2009})}\BibitemShut {NoStop}%
\bibitem [{\citenamefont {Reinhart-King}\ \emph {et~al.}(2005)\citenamefont
  {Reinhart-King}, \citenamefont {Dembo},\ and\ \citenamefont
  {Hammer}}]{reinhart-king_dynamics_2005}%
  \BibitemOpen
  \bibfield  {author} {\bibinfo {author} {\bibfnamefont {C.~A.}\ \bibnamefont
  {Reinhart-King}}, \bibinfo {author} {\bibfnamefont {M.}~\bibnamefont
  {Dembo}}, \ and\ \bibinfo {author} {\bibfnamefont {D.~A.}\ \bibnamefont
  {Hammer}},\ }\href {\doibase 10.1529/biophysj.104.054320} {\bibfield
  {journal} {\bibinfo  {journal} {Biophysical Journal}\ }\textbf {\bibinfo
  {volume} {89}},\ \bibinfo {pages} {676} (\bibinfo {year} {2005})},\ \bibinfo
  {note} {publisher: Elsevier}\BibitemShut {NoStop}%
\bibitem [{\citenamefont {Fournier}\ \emph {et~al.}(2010)\citenamefont
  {Fournier}, \citenamefont {Sauser}, \citenamefont {Ambrosi}, \citenamefont
  {Meister},\ and\ \citenamefont {Verkhovsky}}]{fournier_force_2010}%
  \BibitemOpen
  \bibfield  {author} {\bibinfo {author} {\bibfnamefont {M.~F.}\ \bibnamefont
  {Fournier}}, \bibinfo {author} {\bibfnamefont {R.}~\bibnamefont {Sauser}},
  \bibinfo {author} {\bibfnamefont {D.}~\bibnamefont {Ambrosi}}, \bibinfo
  {author} {\bibfnamefont {J.-J.}\ \bibnamefont {Meister}}, \ and\ \bibinfo
  {author} {\bibfnamefont {A.~B.}\ \bibnamefont {Verkhovsky}},\ }\href
  {\doibase 10.1083/jcb.200906139} {\bibfield  {journal} {\bibinfo  {journal}
  {The Journal of Cell Biology}\ }\textbf {\bibinfo {volume} {188}},\ \bibinfo
  {pages} {287} (\bibinfo {year} {2010})}\BibitemShut {NoStop}%
\bibitem [{\citenamefont {Palecek}\ \emph {et~al.}(1997)\citenamefont
  {Palecek}, \citenamefont {Loftus}, \citenamefont {Ginsberg}, \citenamefont
  {Lauffenburger},\ and\ \citenamefont
  {Horwitz}}]{palecek_integrin-ligand_1997}%
  \BibitemOpen
  \bibfield  {author} {\bibinfo {author} {\bibfnamefont {S.~P.}\ \bibnamefont
  {Palecek}}, \bibinfo {author} {\bibfnamefont {J.~C.}\ \bibnamefont {Loftus}},
  \bibinfo {author} {\bibfnamefont {M.~H.}\ \bibnamefont {Ginsberg}}, \bibinfo
  {author} {\bibfnamefont {D.~A.}\ \bibnamefont {Lauffenburger}}, \ and\
  \bibinfo {author} {\bibfnamefont {A.~F.}\ \bibnamefont {Horwitz}},\ }\href
  {\doibase 10.1038/385537a0} {\bibfield  {journal} {\bibinfo  {journal}
  {Nature}\ }\textbf {\bibinfo {volume} {385}},\ \bibinfo {pages} {537}
  (\bibinfo {year} {1997})},\ \bibinfo {note} {number: 6616 Publisher: Nature
  Publishing Group}\BibitemShut {NoStop}%
\bibitem [{\citenamefont {Sadhanasatish}\ \emph {et~al.}(2023)\citenamefont
  {Sadhanasatish}, \citenamefont {Augustin}, \citenamefont {Windgasse},
  \citenamefont {Chrostek-Grashoff}, \citenamefont {Rief},\ and\ \citenamefont
  {Grashoff}}]{sadhanasatish_molecular_2023}%
  \BibitemOpen
  \bibfield  {author} {\bibinfo {author} {\bibfnamefont {T.}~\bibnamefont
  {Sadhanasatish}}, \bibinfo {author} {\bibfnamefont {K.}~\bibnamefont
  {Augustin}}, \bibinfo {author} {\bibfnamefont {L.}~\bibnamefont {Windgasse}},
  \bibinfo {author} {\bibfnamefont {A.}~\bibnamefont {Chrostek-Grashoff}},
  \bibinfo {author} {\bibfnamefont {M.}~\bibnamefont {Rief}}, \ and\ \bibinfo
  {author} {\bibfnamefont {C.}~\bibnamefont {Grashoff}},\ }\href {\doibase
  10.1126/sciadv.adg3347} {\bibfield  {journal} {\bibinfo  {journal} {Science
  Advances}\ }\textbf {\bibinfo {volume} {9}},\ \bibinfo {pages} {eadg3347}
  (\bibinfo {year} {2023})},\ \bibinfo {note} {publisher: American Association
  for the Advancement of Science}\BibitemShut {NoStop}%
\bibitem [{\citenamefont {Vicente-Manzanares}\ \emph
  {et~al.}(2009)\citenamefont {Vicente-Manzanares}, \citenamefont {Choi},\ and\
  \citenamefont {Horwitz}}]{vicente-manzanares_integrins_2009}%
  \BibitemOpen
  \bibfield  {author} {\bibinfo {author} {\bibfnamefont {M.}~\bibnamefont
  {Vicente-Manzanares}}, \bibinfo {author} {\bibfnamefont {C.~K.}\ \bibnamefont
  {Choi}}, \ and\ \bibinfo {author} {\bibfnamefont {A.~R.}\ \bibnamefont
  {Horwitz}},\ }\href {\doibase 10.1242/jcs.018564} {\bibfield  {journal}
  {\bibinfo  {journal} {Journal of Cell Science}\ }\textbf {\bibinfo {volume}
  {122}},\ \bibinfo {pages} {199} (\bibinfo {year} {2009})}\BibitemShut
  {NoStop}%
\bibitem [{\citenamefont {Goley}\ and\ \citenamefont
  {Welch}(2006)}]{goley_arp23_2006}%
  \BibitemOpen
  \bibfield  {author} {\bibinfo {author} {\bibfnamefont {E.~D.}\ \bibnamefont
  {Goley}}\ and\ \bibinfo {author} {\bibfnamefont {M.~D.}\ \bibnamefont
  {Welch}},\ }\href {\doibase 10.1038/nrm2026} {\bibfield  {journal} {\bibinfo
  {journal} {Nature Reviews Molecular Cell Biology}\ }\textbf {\bibinfo
  {volume} {7}},\ \bibinfo {pages} {713} (\bibinfo {year} {2006})},\ \bibinfo
  {note} {publisher: Nature Publishing Group}\BibitemShut {NoStop}%
\bibitem [{\citenamefont {Sever}\ and\ \citenamefont
  {Schiffer}(2018)}]{sever_actin_2018}%
  \BibitemOpen
  \bibfield  {author} {\bibinfo {author} {\bibfnamefont {S.}~\bibnamefont
  {Sever}}\ and\ \bibinfo {author} {\bibfnamefont {M.}~\bibnamefont
  {Schiffer}},\ }\href {\doibase 10.1016/j.kint.2017.12.028} {\bibfield
  {journal} {\bibinfo  {journal} {Kidney international}\ }\textbf {\bibinfo
  {volume} {93}},\ \bibinfo {pages} {1298} (\bibinfo {year}
  {2018})}\BibitemShut {NoStop}%
\bibitem [{\citenamefont {Marshall}\ \emph {et~al.}(2003)\citenamefont
  {Marshall}, \citenamefont {Long}, \citenamefont {Piper}, \citenamefont
  {Yago}, \citenamefont {McEver},\ and\ \citenamefont
  {Zhu}}]{marshall_direct_2003}%
  \BibitemOpen
  \bibfield  {author} {\bibinfo {author} {\bibfnamefont {B.~T.}\ \bibnamefont
  {Marshall}}, \bibinfo {author} {\bibfnamefont {M.}~\bibnamefont {Long}},
  \bibinfo {author} {\bibfnamefont {J.~W.}\ \bibnamefont {Piper}}, \bibinfo
  {author} {\bibfnamefont {T.}~\bibnamefont {Yago}}, \bibinfo {author}
  {\bibfnamefont {R.~P.}\ \bibnamefont {McEver}}, \ and\ \bibinfo {author}
  {\bibfnamefont {C.}~\bibnamefont {Zhu}},\ }\href {\doibase
  10.1038/nature01605} {\bibfield  {journal} {\bibinfo  {journal} {Nature}\
  }\textbf {\bibinfo {volume} {423}},\ \bibinfo {pages} {190} (\bibinfo {year}
  {2003})},\ \bibinfo {note} {publisher: Nature Publishing Group}\BibitemShut
  {NoStop}%
\bibitem [{\citenamefont {Braeutigam}\ \emph {et~al.}(2022)\citenamefont
  {Braeutigam}, \citenamefont {Simsek}, \citenamefont {Gompper},\ and\
  \citenamefont {Sabass}}]{braeutigam2022generic}%
  \BibitemOpen
  \bibfield  {author} {\bibinfo {author} {\bibfnamefont {A.}~\bibnamefont
  {Braeutigam}}, \bibinfo {author} {\bibfnamefont {A.~N.}\ \bibnamefont
  {Simsek}}, \bibinfo {author} {\bibfnamefont {G.}~\bibnamefont {Gompper}}, \
  and\ \bibinfo {author} {\bibfnamefont {B.}~\bibnamefont {Sabass}},\
  }\href@noop {} {\bibfield  {journal} {\bibinfo  {journal} {Nature
  Communications}\ }\textbf {\bibinfo {volume} {13}},\ \bibinfo {pages} {2197}
  (\bibinfo {year} {2022})}\BibitemShut {NoStop}%
\bibitem [{\citenamefont {Blanch-Mercader}\ and\ \citenamefont
  {Casademunt}(2013)}]{PhysRevLett.110.078102}%
  \BibitemOpen
  \bibfield  {author} {\bibinfo {author} {\bibfnamefont {C.}~\bibnamefont
  {Blanch-Mercader}}\ and\ \bibinfo {author} {\bibfnamefont {J.}~\bibnamefont
  {Casademunt}},\ }\href {\doibase 10.1103/PhysRevLett.110.078102} {\bibfield
  {journal} {\bibinfo  {journal} {Phys. Rev. Lett.}\ }\textbf {\bibinfo
  {volume} {110}},\ \bibinfo {pages} {078102} (\bibinfo {year}
  {2013})}\BibitemShut {NoStop}%
\bibitem [{\citenamefont {Reeves}\ \emph {et~al.}(2018)\citenamefont {Reeves},
  \citenamefont {Winkler}, \citenamefont {Ziebert},\ and\ \citenamefont
  {Aranson}}]{reeves_rotating_2018}%
  \BibitemOpen
  \bibfield  {author} {\bibinfo {author} {\bibfnamefont {C.}~\bibnamefont
  {Reeves}}, \bibinfo {author} {\bibfnamefont {B.}~\bibnamefont {Winkler}},
  \bibinfo {author} {\bibfnamefont {F.}~\bibnamefont {Ziebert}}, \ and\
  \bibinfo {author} {\bibfnamefont {I.~S.}\ \bibnamefont {Aranson}},\ }\href
  {\doibase 10.1038/s42005-018-0075-7} {\bibfield  {journal} {\bibinfo
  {journal} {Communications Physics}\ }\textbf {\bibinfo {volume} {1}},\
  \bibinfo {pages} {1} (\bibinfo {year} {2018})},\ \bibinfo {note} {publisher:
  Nature Publishing Group}\BibitemShut {NoStop}%
\bibitem [{\citenamefont {Safsten}\ \emph {et~al.}(2022)\citenamefont
  {Safsten}, \citenamefont {Rybalko},\ and\ \citenamefont
  {Berlyand}}]{PhysRevE.105.024403}%
  \BibitemOpen
  \bibfield  {author} {\bibinfo {author} {\bibfnamefont {C.~A.}\ \bibnamefont
  {Safsten}}, \bibinfo {author} {\bibfnamefont {V.}~\bibnamefont {Rybalko}}, \
  and\ \bibinfo {author} {\bibfnamefont {L.}~\bibnamefont {Berlyand}},\ }\href
  {\doibase 10.1103/PhysRevE.105.024403} {\bibfield  {journal} {\bibinfo
  {journal} {Phys. Rev. E}\ }\textbf {\bibinfo {volume} {105}},\ \bibinfo
  {pages} {024403} (\bibinfo {year} {2022})}\BibitemShut {NoStop}%
\bibitem [{\citenamefont {Oakes}\ \emph {et~al.}(2017)\citenamefont {Oakes},
  \citenamefont {Wagner}, \citenamefont {Brand}, \citenamefont {Probst},
  \citenamefont {Linke}, \citenamefont {Schwarz}, \citenamefont {Glotzer},\
  and\ \citenamefont {Gardel}}]{oakes_optogenetic_2017}%
  \BibitemOpen
  \bibfield  {author} {\bibinfo {author} {\bibfnamefont {P.~W.}\ \bibnamefont
  {Oakes}}, \bibinfo {author} {\bibfnamefont {E.}~\bibnamefont {Wagner}},
  \bibinfo {author} {\bibfnamefont {C.~A.}\ \bibnamefont {Brand}}, \bibinfo
  {author} {\bibfnamefont {D.}~\bibnamefont {Probst}}, \bibinfo {author}
  {\bibfnamefont {M.}~\bibnamefont {Linke}}, \bibinfo {author} {\bibfnamefont
  {U.~S.}\ \bibnamefont {Schwarz}}, \bibinfo {author} {\bibfnamefont
  {M.}~\bibnamefont {Glotzer}}, \ and\ \bibinfo {author} {\bibfnamefont
  {M.~L.}\ \bibnamefont {Gardel}},\ }\href {\doibase 10.1038/ncomms15817}
  {\bibfield  {journal} {\bibinfo  {journal} {Nature Communications}\ }\textbf
  {\bibinfo {volume} {8}},\ \bibinfo {pages} {15817} (\bibinfo {year}
  {2017})},\ \bibinfo {note} {publisher: Nature Publishing Group}\BibitemShut
  {NoStop}%
\bibitem [{\citenamefont {Lele}\ \emph {et~al.}(2008)\citenamefont {Lele},
  \citenamefont {Thodeti}, \citenamefont {Pendse},\ and\ \citenamefont
  {Ingber}}]{lele_investigating_2008}%
  \BibitemOpen
  \bibfield  {author} {\bibinfo {author} {\bibfnamefont {T.~P.}\ \bibnamefont
  {Lele}}, \bibinfo {author} {\bibfnamefont {C.~K.}\ \bibnamefont {Thodeti}},
  \bibinfo {author} {\bibfnamefont {J.}~\bibnamefont {Pendse}}, \ and\ \bibinfo
  {author} {\bibfnamefont {D.~E.}\ \bibnamefont {Ingber}},\ }\href {\doibase
  10.1016/j.bbrc.2008.02.137} {\bibfield  {journal} {\bibinfo  {journal}
  {Biochemical and Biophysical Research Communications}\ }\textbf {\bibinfo
  {volume} {369}},\ \bibinfo {pages} {929} (\bibinfo {year}
  {2008})}\BibitemShut {NoStop}%
\bibitem [{\citenamefont {Jiang}\ \emph {et~al.}(2003)\citenamefont {Jiang},
  \citenamefont {Giannone}, \citenamefont {Critchley}, \citenamefont
  {Fukumoto},\ and\ \citenamefont {Sheetz}}]{jiang_two-piconewton_2003}%
  \BibitemOpen
  \bibfield  {author} {\bibinfo {author} {\bibfnamefont {G.}~\bibnamefont
  {Jiang}}, \bibinfo {author} {\bibfnamefont {G.}~\bibnamefont {Giannone}},
  \bibinfo {author} {\bibfnamefont {D.~R.}\ \bibnamefont {Critchley}}, \bibinfo
  {author} {\bibfnamefont {E.}~\bibnamefont {Fukumoto}}, \ and\ \bibinfo
  {author} {\bibfnamefont {M.~P.}\ \bibnamefont {Sheetz}},\ }\href {\doibase
  10.1038/nature01805} {\bibfield  {journal} {\bibinfo  {journal} {Nature}\
  }\textbf {\bibinfo {volume} {424}},\ \bibinfo {pages} {334} (\bibinfo {year}
  {2003})},\ \bibinfo {note} {publisher: Nature Publishing Group}\BibitemShut
  {NoStop}%
\bibitem [{\citenamefont {Chen}\ \emph {et~al.}(2023)\citenamefont {Chen},
  \citenamefont {Saintillan},\ and\ \citenamefont
  {Rangamani}}]{chen_interplay_2023}%
  \BibitemOpen
  \bibfield  {author} {\bibinfo {author} {\bibfnamefont {Y.}~\bibnamefont
  {Chen}}, \bibinfo {author} {\bibfnamefont {D.}~\bibnamefont {Saintillan}}, \
  and\ \bibinfo {author} {\bibfnamefont {P.}~\bibnamefont {Rangamani}},\ }\href
  {\doibase 10.1103/PRXLife.1.023007} {\bibfield  {journal} {\bibinfo
  {journal} {PRX Life}\ }\textbf {\bibinfo {volume} {1}},\ \bibinfo {pages}
  {023007} (\bibinfo {year} {2023})},\ \bibinfo {note} {publisher: American
  Physical Society}\BibitemShut {NoStop}%
\end{thebibliography}
%

\end{document}